\documentstyle[astrobib,psfig]{mn-ab}

\def\taucdm{$\tau$CDM}
\def\lcdm3{$\Lambda$CDM.3}
\newcommand{\refgal}{${V_c=220\:\mbox{km s$^{-1}$}}$}

\newcommand{\hmpc}{\mbox{$h^{-1}$ Mpc}}
\newcommand{\hmsun}{\mbox{$h^{-1}$ $M_{\odot}$}}

\newcommand{\kms}{\mbox{km s$^{-1}$}}

\newcommand{\msun}{\mbox{$M_{\odot}$}}
\newcommand{\Zsun}{\mbox{$Z_{\odot}$}}
\newcommand{\kmsmpc}{\mbox{km s$^{-1}$ Mpc$^{-1}$}}

\newcommand{\cobe} {\emph{COBE}\ }

\def\la{\mathrel{\hbox{\rlap{\hbox{\lower4pt\hbox{$\sim$}}}\hbox{$<$}}}}
\def\ga{\mathrel{\hbox{\rlap{\hbox{\lower4pt\hbox{$\sim$}}}\hbox{$>$}}}}

\title[Semi-Analytic Modelling of Galaxy Formation]
      {Semi-Analytic Modelling of Galaxy Formation: The Local Universe} 
\author[Rachel S. Somerville \& Joel R. Primack]
       {Rachel S. Somerville$^{1,2}$ and Joel R. Primack$^2$\\
        $^1$Racah Institute of Physics, The Hebrew University, Jerusalem\\
        $^2$Physics Department, University of California, Santa Cruz}

\begin{document}

\maketitle

\begin{abstract}
Using semi-analytic models of galaxy formation, we investigate galaxy
properties such as the Tully-Fisher relation, the B and K-band luminosity
functions, cold gas contents, sizes, metallicities, and colours, and compare
our results with observations of local galaxies. We investigate several
different recipes for star formation and supernova feedback, including choices
that are similar to the treatment in Kauffmann, White \& Guiderdoni (1993) and
Cole et al. (1994) as well as some new recipes. We obtain good agreement with
all of the key local observations mentioned above. In particular, in our best
models, we simultaneously produce good agreement with both the observed B and
K-band luminosity functions and the I-band Tully-Fisher relation. Improved
cooling and supernova feedback modelling, inclusion of dust extinction, and an
improved Press-Schechter model all contribute to this success. We present
results for several variants of the CDM family of cosmologies, and find that
models with values of $\Omega_0 \simeq 0.3$--$0.5$ give the best agreement with
observations.
\end{abstract}

\begin{keywords}
galaxies: formation -- galaxies: evolution -- cosmology: theory
\end{keywords}

\section{Introduction}
\label{sec:intro}
Over the past decade and a half, a great deal of progress towards a
qualitative understanding of galaxy properties has been made within
the framework of the Cold Dark Matter (CDM) picture of structure
formation (e.g., \citeNP{bfpr}). However, N-body simulations with gas
hydrodynamics still have difficulty reproducing the observed
properties of galaxies in detail (cf. \citeNP{steinmetz:review}). It
is apparent that there must be additional physics that needs to be
included in order to obtain realistic galaxies in the CDM
framework. It is likely that many processes (e.g. cooling, star
formation, supernova feedback, etc.)  form a complicated feedback
loop. It is not computationally feasible to include realistic physics
over the required dynamic range in N-body simulations of significant
volume, especially because we do not currently understand the details
of these processes.

Semi-analytic models (SAMs) of galaxy formation are embedded within
the framework of a CDM-like initial power spectrum and the theory of
the growth and collapse of fluctuations through gravitational
instability. They include a simplified yet physical treatment of gas
cooling, star formation, supernova feedback, and galaxy merging. The
Monte-Carlo approach enables us to study individual objects or global
quantities. Many realizations can be run in a moderate amount of time
on a workstation. Thus SAMs are an efficient way of exploring the
large parameter space occupied by the unknowns associated with star
formation, supernova feedback, the stellar initial mass function,
metallicity yield, dust extinction, etc. However it is not only a
question of computational efficiency: the macroscopic picture afforded
by the semi-analytic method provides an important level of
understanding that would be difficult to achieve by running an N-body
simulation, even if we had an arbitrarily large and fast computer.

The semi-analytic approach to galaxy formation was formulated in
\citeN{wf}, but this approach was not Monte-Carlo based and thus could
only predict average quantities. The Monte-Carlo approach was
primarily developed independently by two main groups, which we shall
refer to as the ``Munich'' group
\cite{kwg,k:faintcounts,k:bo,k:ellip,k:disks,kns,kauffmann:cm} 
and the ``Durham'' group
\cite{cafnz,heyl:95,bcf:96a,bcf:96b,baugh:97}, because the majority of
the members of these groups are associated with the
Max--Planck--Institut f\"{u}r Astrophysik in Garching, near Munich,
Germany, and the University of Durham, U.K., respectively. Similar
models have also been investigated by
\citeN{lacey:91} and \citeN{lacey:93}. This work has shown that it is
possible to reproduce, at least qualitatively, many fundamental
observations in the simple framework of SAMs. These include the galaxy
luminosity function (LF), the Tully-Fisher relation (TFR), the
morphology-density relation, cold gas content as a function of
luminosity and environment, and trends of galaxy colour with
morphology and environment. However, some unsolved puzzles remain. A
fundamental discrepancy has been the inability of the models to
\emph{simultaneously} reproduce the observed Tully-Fisher
relation and the B-band luminosity function in any CDM-type
cosmology \cite{kwg,cafnz,heyl:95}. Another problem, thought to be
generic to the hierarchical structure formation scenario, is the
tendency of larger (more luminous) galaxies to have bluer colours than
smaller (less luminous) ones, in contrast to the observed trend. We
shall discuss these and other problems in detail in this
paper\footnote{Naturally these groups have continued to modify and
improve their models. In this paper, when we make general statements
about the published Munich and Durham models, we refer to work that
was published before February 1998}.

This paper has several goals. We describe the ingredients of our
models and show that they reproduce fundamental observations of the
local universe. We repeat the calculations of several quantities that
have been studied before using SAMs, and one might wonder why this is
worthwhile. First, this will serve as a reference point for future
papers in which we will use these models to study new
problems. Second, the previous studies have been spread out over
several years with different quantities being presented in different
papers. Over this time the models themselves have evolved. We
therefore think it will be useful to have all of these results
presented in the same place in a homogeneous manner. In addition, the
two main groups have not always studied the same quantities, and when
they have, they have not always presented their results in a way that
is directly comparable. This makes it difficult for the non-expert to
judge just how different these two approaches really are. Moreover,
because the models differ in so many details, it is impossible to
determine which particular ingredients are responsible for certain
differences in the results. Two of the important differences that we
are particularly interested in are the parameterization of star
formation and supernova feedback. We shall investigate the results of
varying these recipes while keeping the other ingredients fixed. We
also include some physical effects that have previously been
neglected, and show that some of the problems that have plagued
previous models can be alleviated. We investigate the importance of
the underlying cosmology by examining the same quantities in a wide
range of different cosmologies, spanning currently popular variants of
the Cold Dark Matter (CDM) family of models.

The structure of the paper is as follows. In Section~\ref{sec:models}
we describe the basic physical ingredients of the models and briefly
summarize the SAM approach.  In Section~\ref{sec:models:params}, we
summarize the model parameters and describe how we set them. In
Section~\ref{sec:varyparam}, we illustrate the effects of varying the
free parameters and the star formation and supernova feedback
recipes, using the properties of galaxies within a ``Local Group''
sized halo as an illustration. In Section~\ref{sec:results}, we
present the results of our models for fundamental global quantities
and galaxy properties, illustrate the effects of different choices of
star formation and supernova feedback recipes on these quantities, and
compare our results with previous work. We summarize and discuss our
results in Section~\ref{sec:summary}.

\section{Basic Ingredients}
\label{sec:models}
In this section we summarize the simplified but physical treatments of
the basic physics used in our SAMs. This includes the growth of
structure in the dark matter component, shock heating and radiative
cooling of hot gas in virialized dark matter halos, the formation of
stars from the cooled gas, the reheating of cold gas by supernova
feedback, the evolution of the stellar populations, and mergers of
galaxies within the dark matter halos. There are many assumptions
implicit in this modelling and in addition to describing the choices
we have adopted in our fiducial models, we also remark upon some
relevant details of the assumptions made in previously published work.

Our models have been developed independently, but very much in the
spirit of \citeN[hereafter KWG93]{kwg},
\citeN[hereafter CAFNZ94]{cafnz}, and subsequent work by these
groups. We refer the reader to this literature for a more
detailed introduction to the SAM approach, which here is summarized
rather briefly. A more detailed review of the literature and
description of an earlier version of our models is given in
\citeN{mythesis}.

\subsection{Cosmology}
\label{sec:models:cosmo}
Most of the previous SAM work has been in the context of standard cold
dark matter (SCDM), $\Omega_0=1$, $H_0 = 50\, \kmsmpc$,
$\sigma_8=0.67$. However, this model has now been discredited many
times in many different ways. Particularly relevant to our work is the
problem that this model overproduces objects on galaxy scales relative
to cluster scales. In addition the normalization $\sigma_8=0.67$ is
highly inconsistent with the \cobe data, which requires $\sigma_8 \sim
1.2$ \cite{gorski:96} for this model. Many alternative variants of CDM
have been suggested. We have chosen illustrative examples of popular
variants of the CDM family of models, spanning the observationally
plausible range of parameter space.
\begin{table}
\caption{Parameters of Cosmological Models. From left to right, the tabulated
quantities are: the matter density, the density in the form of a
cosmological constant in units of the critical density, the Hubble
parameter, the baryon density in units of the critical density, the
age of the universe in Gyr, the slope of the primordial power
spectrum, and the linear rms mass variance on a scale of $8\hmpc$.}
\begin{center}
\begin{tabular}{lccccccc}
\hline
Model& $\Omega_0$ & $\Omega_\Lambda$ & $h$ & $\Omega_{b}$ 
& $t_0$ & $n$ & $\sigma_8$ \\
\hline
SCDM & 1.0 & 0.0 & 0.5 & 0.072 & 13.0 & 1.0 & 0.67 \\          
\taucdm & 1.0 & 0.0 & 0.5 & 0.072 & 13.0 & 1.0 & 0.60\\
$\Lambda$CDM.5 & 0.5 & 0.5 & 0.6 & 0.050 & 13.5 & 0.9 & 0.87\\
OCDM.5 & 0.5 & 0.0 & 0.6 & 0.05 & 12.3 & 1.0 & 0.85\\     
$\Lambda$CDM.3 & 0.3 & 0.7 & 0.7 & 0.037 & 13.5 &1.0 & 1.0\\
OCDM.3 & 0.3 & 0.0 & 0.7 & 0.037 & 11.3 & 1.0 & 0.85\\
\hline
\end{tabular}
\end{center}
\label{tab:cosmo}
\end{table}
%
\begin{figure}
\centerline{\psfig{file=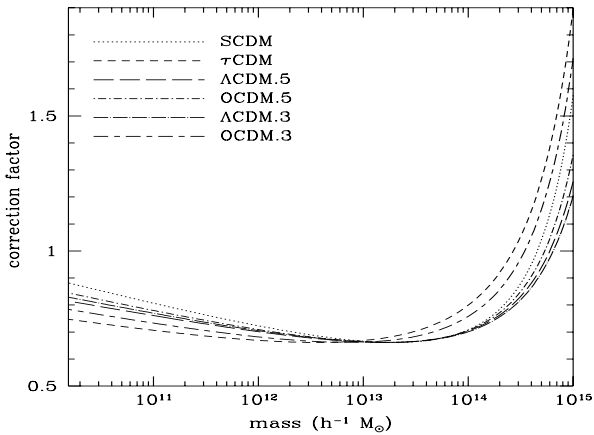,height=6truecm,width=8truecm}}
\caption{The mass function from the improved
Press-Schechter model proposed by
\protect\cite{sheth-tormen} 
divided by the standard Press-Schechter mass function. This
``correction factor'' is a function of redshift as well as halo mass,
and here is shown for $z=0$.}
\label{fig:pscorrec}
\end{figure}
We have retained the standard SCDM model for comparison with previous
work, and consider one other model with $\Omega_0=1$, the $\tau$CDM
model of \citeN{ebw:92}. For our purposes, the properties of this
model are very similar to other popular $\Omega=1$ models such as
tilted CDM ($n \sim 0.8$) and models with an admixture of hot dark
matter (CHDM; \citeNP{primack:95}). We also consider open
($\Omega_{\Lambda}=0$) and flat $\Omega_0+\Omega_{\Lambda}=1$) models
with $\Omega_0=0.3$ and $\Omega_0=0.5$. We have assumed a Hubble
parameter $h=0.7$ for the $\Omega_0=0.3$ models, $h=0.6$ for the
$\Omega_0=0.5$ models, and $h=0.5$ for the $\Omega_0=1$ models ($H_0
\equiv 100 h \kmsmpc$). For the $\Omega_0=0.5$ flat model, we have
included a mild tilt ($n=0.9$) to better simultaneously fit the power
on \cobe and cluster scales. For all the models, in computing the
power spectrum we have assumed the baryon fraction implied by the
observations of
\citeN{tytler:99}, $\Omega_{\rm b}=0.019\, {\rm h^{-2}}$. All models assume
$T/S=0$ (no contribution from tensor modes). We use the fitting
functions of \citeN{bw:97}, modified to account for the presence of
baryons using the prescription of \citeN{hu:1996}, to obtain the
linear power spectra and \cobe normalizations. The normalization
$\sigma_8$ is roughly consistent with the $z=0$ cluster abundance and
the \cobe measurement except in the case of SCDM and the
$\Omega_0=0.3$ open model, for which we have used the cluster
normalization. The parameters of the cosmologies are presented in
Table~\ref{tab:cosmo}.

\subsection{Dark Matter Merger Trees}
\label{sec:models:trees}
The extended Press-Schechter formalism \cite{bower:91,bcek,lc:93}
provides us with an expression for the probability that a halo of a
given mass $m_0$ at redshift $z_0$ has a progenitor of mass $m_1$ at
some larger redshift $z_1$. Several methods of creating Monte-Carlo
realizations of the merging histories of dark matter halos (``merger
trees'') using this formalism have been developed
\cite{kw,block2,sk}. Although the agreement of the Press-Schechter
model with N-body simulations is in some ways surprisingly good given
the simplifications involved, recent work has emphasized that there
are non-negligible discrepancies. Several authors
\cite{gsphk:98,slkd,tormen:98,tozzi:97} have now
reported the same results using different N-body codes and different
methods of identifying halos, indicating that the problems are
unlikely to be explained by numerical effects. \citeN{gsphk:98} showed
that for a wide variety of CDM-type models, the $z=0$ Press-Schechter
mass function agrees well with simulations on mass scales $M \ga
10^{14} \msun$, but on smaller scales the Press-Schechter theory
over-predicts the number of halos by about a factor of 1.5 to 2. The
precise factor varies somewhat depending on the cosmological model and
power spectrum and the way the Press-Schechter model is
implemented. However, this problem cannot be solved by adjusting the
critical density for collapse, $\delta_{c,0}$. In addition, the
Press-Schechter model predicts stronger evolution with redshift in the
halo mass function than is observed in the simulations
\cite{mike-thesis,slkd}. \citeN{tormen:98} finds a similar
behaviour when comparing the predictions of the extended
Press-Schechter theory with the conditional mass function of
cluster-sized halos in simulations.  \citeN{slkd} also compared the
extended Press-Schechter model with the results of dissipationless
N-body simulations, and investigated how well the distribution of
progenitor number and mass in the simulations agrees with that
produced by the merger-tree method of \citeN{sk}. They found that the
distributions of progenitor number and mass obtained in the merging
trees, which have been deliberately constructed to reproduce the
Press-Schechter model, are skewed towards larger numbers of smaller
mass progenitors than are found in the simulations. This problem is
endemic to any method based on the extended Press-Schechter
model. However, the \emph{relative} properties of progenitors within a
halo of a given mass are very similar in the merger trees and the
simulations. This suggests that the merger trees should provide a
fairly reliable framework for modelling galaxy formation, if the
overall error in the Press-Schechter mass function is corrected
for. An improved version of the Press-Schechter model, which gives
good agreement with simulations for a variety of cosmologies, has
recently been proposed by \citeN{sheth-tormen}. The ``correction
factor'', i.e. the mass function from the Sheth-Tormen model divided
by the standard Press-Schechter mass function, is shown in
Fig.~\ref{fig:pscorrec}.

In the merger-tree method of \citeN{sk}, used here, the merging
history of a dark matter halo is constructed by sampling the paths of
individual particle trajectories using the excursion set formalism
\cite{bcek,lc:93}. It does not require the imposition of a grid in
mass or redshift, nor are merger events required to be binary. The
redshifts of branching events (i.e. halo mergers) and the masses of
the progenitor halos at each stage are chosen randomly using
Monte-Carlo techniques, such that the overall distribution satisfies
the average predicted by the extended Press-Schechter theory. Thus
when we subsequently refer to a ``realization'' we mean a particular
Monte-Carlo realization of the halo merging history. This is the most
important stochastic ingredient in the models. In order to make the
tree finite, it is necessary to impose a minimum mass $m_{\rm
min}$. Although the contribution of mass from halos smaller than
$m_{\rm min}$ is included, we do not trace the merging history of
halos with masses less than $m_{\rm min}$, but rather assume that this
mass is accreted as a diffuse component. Here we take $m_{\rm min}$ to
be equal to the mass corresponding to a halo with a circular velocity
of 40 km/s at the relevant redshift. We argue that galaxies are
unlikely to form in halos smaller than this because the gas will be
photoionized and unable to cool
\cite{weinberg:97,forcada-miro:uv}. 

For the prediction of global quantities, we run a grid of halo masses
(typically $\sim50$ halos from $10\, m_{\rm min}$ to $V_c=1500$ km/s),
and weight the results with the overall number density for the
appropriate mass and redshift, using the improved Press-Schecter model
of \citeN{sheth-tormen}. We run many such grids and average the
results.

\subsection{Gas Cooling}
\label{sec:models:cooling}
\begin{figure}
\centerline{\psfig{file=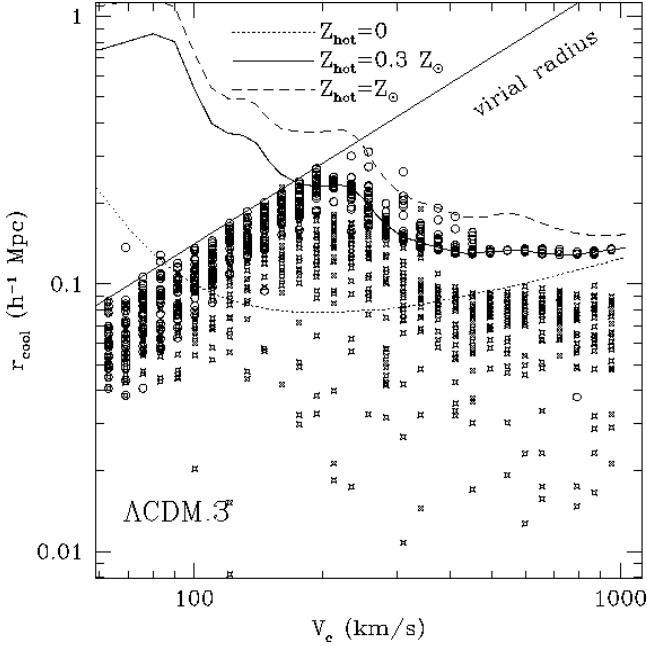,height=8.5truecm,width=8.5truecm}}
\caption{Cooling radius of halos as a function of circular velocity. The
straight diagonal line shows the virial radius, which the cooling radius may
not exceed. The curved lines show the cooling radius predicted by the literal
static halo cooling model (see text), assuming that the hot gas has primordial,
0.3 solar, or solar metallicity. Open circles show the application of the
static halo model within the merger trees, and crosses show the dynamic halo
model (see text), assuming a fixed metallicity of 0.3 solar. Earlier conversion
of gas from the hot to cold phase and reheating of hot gas by halo mergers
results in a lower cooling efficiency for large halos in the dynamic halo
model.}
\label{fig:cooling}
\end{figure}
\subsubsection{Cooling in Static Halos}
Gas cooling is modelled using an approach similar to the one
introduced by \citeN{wf}. We assume that each newly formed halo, at
the top level of the tree, contains pristine hot gas that has been
shock heated to the virial temperature of the halo ($T$) and that the
gas traces the dark matter. The rate of specific energy loss due to
radiative cooling is given by the cooling function $\Lambda(T)$. We
can then derive an expression for the critical density which will
enable the gas to cool within a timescale $\tau_{\rm cool}$:
\begin{equation}
\rho_{\rm cool} = \frac{3}{2} \frac{\mu m_p}{\chi_e^2} 
\frac{k_{B}T}{\tau_{\rm cool} \Lambda(T)}
\end{equation}
where $\mu m_p$ is the mean molecular weight of the gas and $\chi_e
\equiv n_e/n_{\rm tot}$ is the number of electrons per particle. 
 Assuming that
the gas is fully ionized and has a helium fraction by mass $Y=0.25$,
\begin{equation}
\rho_{\rm cool} = 3.52 \times 10^{7} 
\frac{k_{B}T}{\tau_{\rm cool} \Lambda_{23}(T)},
\end{equation}
where $k_{B}T$ is in degrees Kelvin, $\tau_{\rm cool}$ is in Gyr, and
$\Lambda_{23}(T) \equiv \Lambda(T)/(10^{-23}$ ergs s$^{-1}$ cm$^3$). The virial
temperature is approximated as $k_{B}T= 71.8 \sigma_{\rm vir}^2$, where
$\sigma_{\rm vir}$ is the virial velocity dispersion of the halo. If we assume
a form for the gas density profile $\rho_g(r)$, we can now invert this
expression to obtain the ``cooling radius'', defined as the radius within which
the gas has had time to cool within the timescale $\tau_{\rm cool}$. For the
simplified choice of the singular isothermal sphere, this gives
\begin{equation}
r_{\rm cool} = \left(\frac{\rho_0}{\rho_{\rm cool}}\right)^{1/2}
\end{equation}
where $\rho_0 = f_{\rm hot}V^2_{c}/(4\pi G)$, $f_{\rm hot}$ is the hot gas
fraction in the cooling front and $V_c = \sqrt{2} \sigma_{\rm vir}$ is the
circular velocity of the halo.

We use the cooling function of \citeN{sutherland:93}. The cooling
function $\Lambda(T)$ is also metallicity dependent. In this paper, we
assume that the hot gas has an average [Fe/H] = 0.3 $\Zsun$ at all
redshifts and for all halo masses. This value is typical of the hot
gas in clusters from $z=0$ to $z\sim 0.3$ \cite{mushotzky:97}. We will
treat chemical enrichment in more detail and investigate the effects
on cooling in a future paper. In practice, we find that the results at
$z=0$ are very similar regardless of whether we use the
self-consistently modelled hot gas metallicity in the cooling
function, or a fixed metallicity [Fe/H] = 0.3 $\Zsun$.

We divide the time interval between halo mergers (branchings) into
small time-steps. For a time-step $\Delta t$, the cooling radius
increases by an amount $\Delta r$ and we assume that the mass of gas
that cools in this time-step is ${\rm d}m_{\rm cool} = 4 \pi r^2_{\rm
cool} \rho_g(r_{\rm cool}) \Delta r$. The cooling radius is not
allowed to exceed the virial radius, and the amount of gas that can
cool in a given timestep is not allowed to exceed the hot gas
contained within the virial radius of the halo. For small halos, and
at high redshift, the cooling is therefore effectively limited by the
accretion rate. New hot gas is constantly accreted as the halo
grows. When we construct the merging tree, we keep track of the amount
of diffuse mass (i.e. halos below the minimal mass $m_{\rm min}$)
accreted at every branching, $m_{\rm acc}$. The mass of hot gas
accreted between branchings is then $f_{\rm bar}\, m_{\rm acc}$, where
$f_{\rm bar}
\equiv \Omega_b/\Omega_0$ is the universal baryon fraction. We
assume that the mass accretion rate is constant over the time interval
between branchings, which is what one would expect from the spherical
collapse model (see Appendix). We also require that even if the gas is
able to cool, it falls onto the disk at a rate given by the sound
speed of the gas, $c_s = (5k_B T/3\mu m_p)^{1/2} \sim 1.3
\sigma_v$, where $\sigma_v$ is the 1-D velocity dispersion of the
halo. Note that $c_s$ is approximately equal to the dynamical velocity
of the halo, and that N-body simulations with hydrodynamics and
cooling show that the radial infall velocity of cooling gas within the
virial radius is generally close to this value \cite{evrard:96}.

\subsubsection{Cooling and Heating in Merging Halos}
\label{sec:models:cooling:mergers}
In the simplest version of this approach to modelling gas cooling in
dark matter halos, we imagine that a halo of a given circular velocity
$V_c$, with a corresponding virial temperature $T$, forms at time
$t=0$ and grows isothermally, gradually cooling at the rate given by
the cooling function $\Lambda(T)$ as described above. In this model,
the cooling time $\tau_{\rm cool}$ is the age of the Universe at the
current redshift, and the gas fraction in the cooling front $f_{\rm
hot}$ is always equal to the universal baryon fraction $f_{\rm
bar}$. We refer to this picture as the ``static halo'' cooling model,
as it does not account for the dynamical effects of halo mergers.

However, in the hierarchical framework of the merger trees, most halos
are built up from merging halos that have experienced cooling, star
formation, and feedback in previous time-steps. This modifies the gas
fraction $f_{\rm hot}$ in the cooling front. The virial velocity and
temperature change discontinuously following a merger, and merger
events may shock heat the cooling gas. We have developed a ``dynamic
halo'' cooling model that incorporates these effects in the following
way.

For top-level halos (halos with all progenitors smaller than the
minimal mass $m_{\rm min}$), the gas fraction $f_{\rm hot}$ is assumed
to be equal to the universal baryon fraction $f_{\rm bar}$, and the
cooling time $\tau_{\rm cool}$ is the time elapsed since the initial
collapse of the halo. Subsequently, when a halo forms from the merging
of two or more halos larger than $m_{\rm min}$, we determine whether
the mass of the largest progenitor $m_1$ comprises more than a
fraction $f_{\rm reheat}$ of the post-merger mass $m_0$. If so, the
cooling radius and cooling time of the new halo are set equal to those
of the largest progenitor. The gas fraction in the cooling front is
taken to be $f_{\rm hot} = m_{\rm hot}/m_{\rm tot}(>r_{\rm cool})$,
where $m_{\rm hot}$ is the sum of the hot gas masses of all the
progenitors, and $m_{\rm tot}(r>r_{\rm cool})$ is the total mass
contained between the cooling radius and the virial radius of the halo
(assuming an isothermal profile). If $m_1/m_0 < f_{\rm reheat}$, we
assume that the hot gas within all the progenitor halos is reheated to
the virial temperature of the new halo, and the cooling radius and
cooling time are reset to zero. The gas fraction in the cooling front
is then $f_{\rm hot}=m_{\rm hot}/m_{0}$. Note that $f_{\rm hot}$ may
in principle be larger than $f_{\rm bar}$ due to reheating by
supernova, but in general $f_{\rm hot} < f_{\rm bar}$ because of
previous gas cooling and consumption. 

We can apply the main simplifying assumptions of the static halo
cooling model within the merging trees, i.e. we always assume $f_{\rm
hot} =f_{\rm bar}$ and $\tau_{\rm cool}$ equal to the age of the
Universe at any given time, and do not reheat the gas after any halo
mergers. The results differ somewhat from the literal static halo
model because we do not allow the cooling rate to exceed the available
supply of hot gas, or to exceed the sound speed constraint, and
because the progenitor halos cool at different
temperatures. Fig.~\ref{fig:cooling} shows the cooling radius in the
literal static cooling model, and in the application of the static
cooling model within the merging trees. For low $V_c$ halos, the
cooling is limited by the available collapsed gas supply (i.e. $r_{\rm
cool} > r_{\rm vir}$). For larger halos the results are similar to the
prediction of the literal static halo model. However, the dynamic halo
model (crosses) predicts significantly less cooling in large halos,
due to the lower values of $f_{\rm hot}$ and the reheating by halo
mergers. Note that the cooling model used by the Munich group more
closely resembles the ``static halo'' model, and the cooling model
used by the Durham group is more similar to our ``dynamic halo''
cooling model\footnote{Note that in earlier versions of our models
(e.g. \citeNP{mythesis}), as in the Munich models, we prevented gas
from cooling altogether in large halos by applying an arbitrary
cutoff. We no longer apply this cutoff.} In this paper, we will show
results for both cooling models.

\subsection{Disk Sizes}
\label{sec:models:sizes}
To obtain a very rough estimate of the sizes of disks that form in our
models, we adopt the general picture of \citeN{fall-efstathiou}, in
which the gas collapse is halted by angular momentum conservation. We
define $\lambda_{H}$ to be the dimensionless spin parameter of the
halo, $\lambda_{H} \equiv J|E|^{1/2}G^{-1}M^{-5/2}$, where $J$ is the
angular momentum, $M$ is the mass and $E$ is the energy of the dark
matter halo. We assume that the gas has the same specific angular
momentum as the dark matter, and collapses to form a disk with an
exponential profile. For a singular isothermal halo, the scale radius
of the disk that forms is then $r_s = 1/\sqrt{2} \lambda_{H} r_i$,
where $r_i$ is the radius before collapse (in our models, $r_i = {\rm
min}(r_{\rm cool}, r_{\rm vir}$). We neglect the modification of the
inner profile of the dark matter due to the infall of the baryons,
which will tend to lead to smaller disks
\cite{blumenthal:86,flores:93,mmw:98}.

The distribution of $\lambda_{H}$ found in N-body simulations
\cite{barnes-efstathiou,kravtsov:98,lemson:97} is a rather broad
log-normal with mean $\langle \lambda_{H} \rangle =0.05$. It is likely
that in order to obtain a realistic distribution of galaxy sizes and
surface brightnesses, we should consider a range of values of
$\lambda_{H}$ as seen in the above simulations
\cite{dss:97,mmw:98}. 
However, it is not known how $\lambda_{H}$ is affected by mergers, so
we do not know how to propagate this quantity through the merging
trees. We should also use a more realistic halo profile than the
singular isothermal sphere. We intend to address the modelling of disk
sizes in more detail in the future. For the present, we use
$\lambda_{H}=0.05$ for all halos.

\begin{figure}
\centerline{\psfig{file=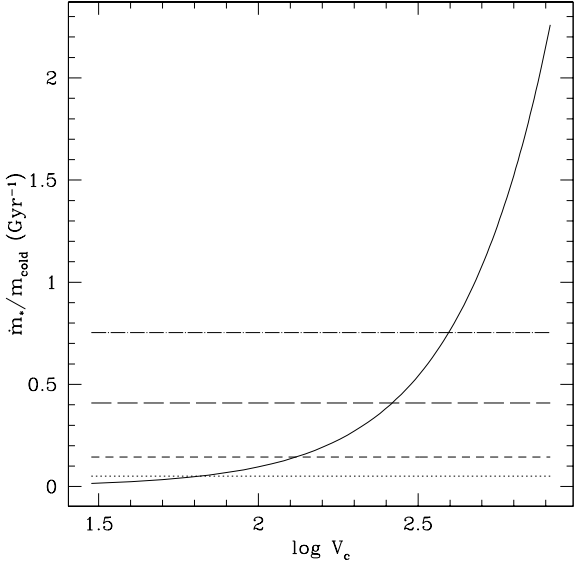,height=8truecm,width=8truecm}}
\caption{The star formation rate per unit mass of cold gas (star
formation efficiency) as a function of circular velocity. The solid
curved line shows the Durham star formation law (SFR-D), which has no
explicit dependence on redshift, and the horizontal dashed lines show
the Munich star formation law (eqn. SFR-M), for $z=5,3,1,0$ from
highest to lowest, respectively.}
\label{fig:sflaws}
\end{figure}
\subsection{Star Formation}
\label{sec:models:sf}
The star formation recipes that we will consider in this paper are of
the general form
\begin{equation}
\dot{m}_{*} = \frac{m_{\rm cold}}{\tau_{*}},
\end{equation} 
where $m_{\rm cold}$ is the total mass of cold gas in the disk and we
hide all of our ignorance in the efficiency factor $\tau_{*}$.  The
simplest possible choice is to assume that $\tau_{*}=\tau_{*}^0$ is
constant. This would imply that once it is cold, gas is converted to
stars with the same efficiency in disks of all sizes and at all
redshifts. We shall refer to this recipe as SFR-C.

Another choice is a power law, in which the star formation efficiency
is a function of the circular velocity of the galaxy:
\begin{eqnarray}
\dot{m}_{*} & = &\frac{m_{\rm cold}}{\tau_{*}(V_c)} \\ 
\tau_{*}(V_c) & = & \tau_{*}^0 \left(\frac{V_c}{V_0}\right)^{\alpha_{*}}
\end{eqnarray}
where $m_{\rm cold}$ is the mass of cold gas in the disk, $\tau_{*}^0$ and
$\alpha_{*}$ are free parameters, and $V_0=300\, \kms$ is an arbitrary
normalization factor. This is equivalent to the approach used by the Durham
group, and we will refer to this recipe as SFR-D.

The other approach, used by the Munich group, assumes that the timescale for
star formation is proportional to the dynamical time of the disk
\begin{equation}
\dot{m}_{*} = \frac{m_{\rm cold}}{\tau_{*}^{0} \, \tau_{\rm dyn}}.
\end{equation}
Here $\tau_{*}^{0}$ is a dimensionless free parameter, and $\tau_{\rm
dyn}$ is the dynamical time of the galactic disk, $\tau_{\rm dyn} =
r_{\rm disk}/V_c$. Following KWG93, we take $r_{\rm disk}$ to be equal
to one tenth the virial radius of the dark matter halo, and $V_c$ to
be the circular velocity of the halo at the virial radius. For
satellite galaxies, the dynamical time remains fixed at the value it
had when the galaxy was last a central galaxy. We will refer to this
star formation recipe as SFR-M.

It is worth noting the differences in these assumptions and the
implications for the models. The dynamical time $\tau_{\rm dyn}$ at a
given redshift is nearly independent of the galaxy circular
velocity. This is because the spherical collapse model predicts that
the virial radius scales like $r_{\rm vir} \propto V_c$ (see
Appendix). However, the virial radius of a halo with a given circular
velocity increases with time ($r_{\rm vir} \propto (1+z)^{-3/2}$ for
an Einstein-de Sitter universe). This means that SFR-M is
approximately constant over circular velocity but has a higher
efficiency at earlier times (higher redshift). In contrast, SFR-D has
no explicit dependence on redshift but does depend fairly strongly on
the galaxy circular velocity ($\alpha_{*}=-1.5$ in the fiducial Durham
models), so that star formation is less efficient in halos with small
$V_c$. This is illustrated in Fig.~\ref{fig:sflaws}. Because a
``typical'' halo at high redshift is less massive and hence has a
smaller circular velocity in hierarchical models, this has the effect
of delaying star formation until a later redshift, when larger disks
start to form. SFR-M therefore leads to more early star formation. We
will discuss this further in Section~\ref{sec:varyparam}.

\subsection{Supernova Feedback}
\label{sec:models:feedback}
\subsubsection{Previous Feedback Recipes}
\label{sec:models:feedback:previous}
In the Munich and Durham models, the rate of reheating of cold gas is
given by
\begin{equation}
\label{eqn:fb_pl}
\dot{m}_{\rm rh} = \epsilon^0_{\rm SN} (V_c/V_{0})^{-\alpha_{\rm rh}} 
\dot{m}_{*}
\end{equation} 
where $\epsilon^0_{\rm SN}$ and $\alpha_{\rm rh}$ are free parameters,
$\dot{m}_{*}$ is the star formation rate, and $V_0$ is a scaling
factor chosen so that $\epsilon^0_{\rm SN}$ is of order unity
($V_0=400\,\kms$ for the Munich models and 140 \kms for the Durham
models). The Munich group assumes $\alpha_{\rm rh}=2$, whereas the
fiducial models of the Durham group assume a considerably stronger
dependence on circular velocity, $\alpha_{\rm rh} = 5.5$. 

The ``reheated'' gas is removed from the cold gas reservoir. An
important issue is whether the reheated gas remains in the halo in the
form of hot gas, where it will generally cool again on a short
timescale, or is expelled from the potential well of the halo
entirely. In the Munich models (previous to 1998), all the reheated
gas is retained in the halo (G. Kauffmann, private communication). In
the Durham models, all the reheated gas is ejected from the halo. This
gas is then returned to the hot gas reservoir of the halo after the
mass of the halo has doubled (Durham group, private communication). We
find that the results of the models are quite sensitive to whether the
gas is retained in or ejected from the halo. We would therefore like
to find a simple but physical way of modelling the ejection of the
reheated gas from the disk and the halo without introducing an
additional free parameter. To this end, we have introduced the
following modified treatment of supernova feedback.

\subsubsection{The Disk-Halo Feedback Model}
\label{sec:models:feedback:disk-halo}
We assume that the mass profile of the disk is exponential. The
potential energy of an exponential disk with scale radius $r_s$ and
central surface density $\Sigma_0$ is approximately $W \simeq -11.6 G
\Sigma_0^2r_s^3$ \cite{BT}. We can then calculate the {\it rms} escape
velocity for the disk, $\langle v^2_{\rm esc,\, disk} \rangle^{1/2} =
\sqrt{-4W/m_d}$, where $m_d$ is the mass of the disk. Similarly, for
the halo, the {\it rms} escape velocity is $\langle v^2_{\rm esc,\,
halo} \rangle^{1/2} = \sqrt{-4W/m_{halo}} = \sqrt{2}V_c$, using the
virial theorem. As before, we have the free parameter $\epsilon^0_{\rm
SN}$, which we interpret loosely as the fraction of the supernova
energy transferred to the gas in the form of kinetic energy. The rate
at which kinetic energy is transferred to the gas is now
$\dot{\epsilon}_{\rm SN} = \epsilon^0_{\rm SN}\, E_{\rm SN}
\, \eta_{\rm SN}\, \dot{m}_{*}$, where $E_{\rm SN}=10^{51}$ ergs is the
total (kinetic and thermal) energy per supernova, $\eta_{\rm SN}$ is
the number of supernova per solar mass of stars ($\eta_{\rm
SN}=3.2\times10^{-3}$ for the Scalo IMF used here \cite{bc:93}), and
$\dot{m}_{*}$ is the star formation rate. Following the general
arguments of \citeN{dekel-silk}, we now calculate the rate at which
gas can escape from the disk:
\begin{equation}
\dot{m}_{\rm rh,\, disk} = 2 f_{\rm rh, disk} 
	\frac{\dot{\epsilon}_{\rm SN}}{\langle v^2_{\rm esc,\, disk}
\rangle},
\end{equation} 
and from the halo (same expression with $\langle v^2_{\rm esc,\, halo}
\rangle$). The factor $f_{\rm rh, disk}$ is a fudge factor that we 
leave fixed to $f_{\rm rh, disk}=2$. In general, ${\rm d}m_{\rm rh,\,
disk}$ is then larger than ${\rm d}m_{\rm rh,\, halo}$. The gas that
can escape from the disk but not the halo is added to the hot gas in
the halo. Gas that can escape from both the disk and the halo is
removed from the halo entirely. Because of the uncertainties involved,
we do not attempt to model the recollapse of this gas at later times,
so this gas will never be re-incorporated into any halo and is
``lost'' forever.

\subsection{Chemical Evolution}
\label{sec:models:chemev}
We trace chemical evolution by assuming a constant ``effective
yield'', or mean mass of metals produced per mass of stars. The value
of the effective yield, $y$, is treated as a free parameter. We assume
that newly produced metals are deposited in the cold
gas. Subsequently, the metals may be ejected from the disk and mixed
with the hot halo gas, or ejected from the halo, in the same
proportion as the reheated gas, according to the feedback model
described above. The metallicity of each batch of new stars equals the
metallicity of the cold gas at that moment. Note that because enriched
gas may be ejected from the halo, and primordial gas is constantly
being accreted by the halo, this approach is not equivalent to a
standard ``closed box'' model of chemical evolution. Also note that
although we track the metallicity of the hot gas by this procedure, in
this paper we do not use this metallicity to compute the gas cooling
rate (see Section~\ref{sec:models:cooling}).

\subsection{Galaxy Merging}
\label{sec:models:mergers}
\subsubsection{Dynamical Friction and Tidal Stripping}
\label{sec:models:mergers:df}
When halos merge, we assume that the galaxies within them remain
distinct for some time. In this way we eventually end up with many
galaxies within a common dark matter halo, as in groups and
clusters. The central galaxy of the largest progenitor halo becomes
the new central galaxy and the other galaxies become
``satellites''. Following a halo merger event, we assume that the
satellites of the largest progenitor halo remain undisturbed and place
the central galaxies of the other progenitors at a distance $f_{\rm
mrg} r_{\rm vir}$ from the central galaxy, where $f_{\rm mrg}$ is a
free parameter and $r_{\rm vir}$ is the virial radius of the new
parent halo. Satellites of the other progenitors are distributed
randomly around their previous central galaxy, preserving their
relative distance from that galaxy. All the satellites lose energy due
to dynamical friction against the dark matter background and fall in
towards the new central object.

The differential equation for the distance of the satellite from the
center of the halo ($r_{\rm fric}$) as a function of time is given by
\begin{equation}
r_{\rm fric} \frac{{\rm d}r_{\rm fric}}{{\rm d}t} = -0.428 f(\epsilon) 
\frac{Gm_{\rm sat}}{V_c} \ln \Lambda
\end{equation}
\cite{BT,nfw:95}. 
In this expression, $m_{\rm sat}$ is the combined mass of the
satellite's gas, stars, and dark matter halo, and $V_c$ is the
circular velocity of the parent halo. Not to be confused with at least
two other quantities in this paper denoted by the same symbol, here
$\ln \Lambda$ is the Coulomb logarithm, which we approximate as $\ln
\Lambda \approx \ln (1 + m_{h}^2/m_{\rm sat}^2$), where $m_{h}$ is the
mass of the parent halo. The circularity parameter $\epsilon$ is
defined as the ratio of the angular momentum of the satellite to that
of a circular orbit with the same energy: $\epsilon =
J/J_c(E)$. \citeN{lc:93} show that the approximation
$f(\epsilon)=\epsilon^{0.78}$ is a good approximation for $\epsilon >
0.02$. We draw $\epsilon$ for each satellite from a uniform
distribution from 0.02 to 1. A new value of $\epsilon$ is chosen if
the parent halo merges with a larger halo.

As the satellite falls in, its dark matter halo is tidally stripped by
the background potential of the parent halo. We approximate the tidal
radius $r_t$ of the satellite halo by the condition $\rho_{\rm
sat}(r_t) = \rho_{\rm halo}(r_{\rm fric})$, i.e. the density at the
tidal radius equals the density of the background halo at the
satellite's current radial position within the larger halo. The mass
of the satellite halo can then be estimated as the mass within
$r_t$. We assume that both halos can be represented by singular
isothermal spheres, $\rho \propto r^{-2}$. When $r_{\rm fric}$ is less
than or equal to the radius of the central galaxy, the satellite
merges with the central galaxy.

\subsubsection{Satellite-Satellite Mergers}
\label{sec:models:mergers:coll}
Satellite galaxies may also collide with each other as they orbit
within the halo. They may merge or only experience a perturbation
depending on their relative velocities and internal velocity
dispersions. From a simple mean free path argument, one expects
satellites to collide on a time scale
\begin{equation}
\tau_{\rm coll} \sim \frac{1}{\bar{n}\sigma v}
\end{equation}
where $\bar{n}$ is the mean density of galaxies, $\sigma$ is the
effective cross section for a single galaxy, and $v$ is a
characteristic velocity.  High-resolution N-body simulations by
\citeN{makino-hut} indicate that this simple scaling actually
describes the merger rate quite accurately for collisions of galaxy
pairs over a broad range of parameter space. They generalize their
results to obtain an expression for the average time between
collisions in a halo containing $N$ equal mass galaxies:
\begin{eqnarray}
\label{eqn:taucoll}
\lefteqn{\tau_{\rm coll} = 
500\, N^{-2} \left(\frac{r_{\rm halo}}{\rm Mpc}\right)^3
\left(\frac{r_{\rm gal}}{\rm 0.12 \: Mpc}\right)^{-2}} \nonumber
\hspace{1.5truecm}\\ && \left(\frac{\sigma_{\rm gal}}{\rm 100\: km\,
s^{-1}}\right)^{-4}
\left(\frac{\sigma_{\rm halo}}{\rm 300\:km\, s^{-1} }\right)^3 {\rm Gyr}.
\end{eqnarray}
Here $r_{\rm halo}$ is the virial radius of the parent halo, $r_{\rm
gal}$ is the tidal radius of the dark matter bound to the satellite
galaxy, $\sigma_{\rm gal}$ is the internal 1-D velocity dispersion of
the galaxy, and $\sigma_{\rm halo}$ is the 1-D velocity dispersion of
the parent halo. Although this expression was derived for equal mass
galaxies, we use it to assign a collision timescale $t_{\rm coll}$ to
each galaxy using the mass and tidal radius of each individual
sub-halo. The probability that a galaxy will merge in a given timestep
$\Delta t$ is then $P_{\rm mrg} = \Delta t/t_{\rm coll}$. A new
velocity dispersion and mass is assigned to the post-merger sub-halo
by assuming that energy is conserved in the collision, and that the
merger product satisfies the virial relation. Note that we do not
allow random collisions between satellite galaxies and central
galaxies, even though they are in principle quite likely, because we
do not know how to model the cross-section for such events.

\subsubsection{Merger-Induced Starbursts}
There is considerable observational and theoretical evidence that
mergers and interactions between galaxies trigger dramatically
enhanced star formation episodes known as starbursts. When two
galaxies merge according to either of the two processes described
above, we assume that the cold gas is converted to stars at the
enhanced rate $e_{\rm burst} m_{\rm cold}/\tau_{\rm dyn}$, where
$m_{\rm cold}$ is the combined cold gas of both galaxies, and
$\tau_{\rm dyn}$ is the dynamical time of the larger galaxy. The burst
efficiency $e_{\rm burst}$ may depend on the mass ratio of the merging
galaxies, and is typically between 0.50 to 1. The mean properties of
galaxies at $z=0$ are quite insensitive to the details of the
treatment of starbursts, although this process turns out to be quite
important for high redshift galaxies. We develop a more detailed
treatment of starbursts, based on simulations with hydrodynamics and
star formation
\cite{mihos:94,mihos:96}, 
and investigate the implications for high redshift galaxies in a
seperate paper \cite{spf}.

\subsubsection{Merger-Driven Morphology}
\label{sec:models:morph}
Simulations of collisions between nearly equal mass spiral galaxies produce
merger remnants that resemble elliptical galaxies. Accretion of a
low-mass satellite by a larger disk will heat and thicken the disk but
not destroy it \cite{barnes-hernquist}. However the line dividing
these cases is fuzzy and depends on many parameters other than the
mass ratio, such as the initial orbit, the relative inclination,
whether the rotation is prograde or retrograde, etc. We introduce a
free parameter, $f_{\rm bulge}$, which determines whether a galaxy
merger leads to formation of a bulge component. If the mass ratio
$m_{S}/m_{B}$ is greater than $f_{\rm bulge}$, then all the stars from
both galaxies are put into a ``bulge'' and the disk is destroyed (here
$m_{S}$ and $m_{B}$ are the baryonic masses of the smaller and bigger
galaxy, i.e. the sum of the cold gas and stellar masses). If
$m_{S}/m_{B} < f_{\rm bulge}$, then the stars from the smaller galaxy
are added to the disk of the larger galaxy. The cold gas reservoirs of
both galaxies are combined. Additional cooling gas may later form a
new disk. The bulge-to-disk ratio at the observation time can then be
used to assign rough morphological types. \citeN{simien-dev} have
correlated the Hubble type and the B luminosity bulge-to-disk
ratio. Using their results, and following KWG93, we categorize
galaxies with $B/D > 1.52$ as ellipticals, $0.68 < B/D < 1.52$ as S0s,
and $B/D < 0.68$ as spirals. Galaxies with no bulge are classified as
irregulars. As shown by KWG93 and \citeN{bcf:96a}, this approach leads
to model galaxies with morphological properties that are in good
agreement with a variety of observations.

\subsection{Stellar Population Synthesis}
\label{sec:models:popsynth}
Stellar population synthesis models provide the Spectral Energy
Distribution ({SED}) of a stellar population of a single age. These
models must assume an Initial Mass Function (IMF) for the stars, which
dictates the fraction of stars created with a given mass. The model
stars are then evolved according to theoretical evolutionary tracks
for stars of a given mass. By keeping track of how many stars of a
given age are created according to our star-formation recipe, we
create synthesized spectra for the composite population. A free
parameter $f^{*}_{\rm lum}$ effectively determines the stellar
mass-to-light ratio; $f^{*}_{\rm lum}$ is defined as the ratio of the
mass in luminous stars to the total stellar mass, $m^{*}_{\rm
lum}/m^{*}_{\rm tot}$. The remainder is assumed to be in the form of
brown dwarfs, planets, etc. We then convolve the synthesized spectra
for each galaxy with the filter response functions appropriate to a
particular set of observations. In this way we obtain colours and
magnitudes that can be directly compared to observations at any
redshift.

Although this approach is satisfying because it results in quantities
that can be compared directly to observations, there are many
uncertainties inherent in this component of the modelling, as is bound
to be the case with such a complicated problem. The IMF is a major
source of uncertainty. The IMF is fairly well determined in our Galaxy
\cite{scalo}, but we know very little about how universal it is or
whether it depends on metallicity or other environmental effects. The
results are somewhat sensitive to the upper and lower mass cutoffs as
well as the slope of the IMF. Then of course there are the
difficulties of modelling the complex physics involved in stellar
evolution. Some of the major sources of uncertainty mentioned by
\citeN{cwb} are opacities, heavy-element mixture, helium content,
convection, diffusion, mass loss, and rotational mixing. Comparing
three sets of models, \citeN{cwb} find only a 0.05 magnitude
dispersion between the models in $B-V$ colour, but a larger
discrepancy of 0.25 magnitudes in $V-K$ colour and a 25\% dispersion
in the mass-to-visual light ratio. However, they also stress that
there are far greater uncertainties involved in the modelling of young
($< 1$ Gyr) stars, especially stars more massive than 2 \msun.
 
There are currently several versions of stellar population models
available. We have used the Bruzual \& Charlot (GISSEL95) models
\cite{bc:93,cwb}. These models are for solar metallicity stars
only. For the results presented in this paper, we have used a Scalo
\cite{scalo} IMF and the standard Johnson filters provided with
GISSEL95.
%
\begin{table*}
\caption{Cooling/Merging Packages}
\begin{center}
\begin{tabular}{lccc}
\hline
name & cooling & merging & starbursts \\
\hline
Classic & static halo & dynamical friction only & major mergers only;
$e_{\rm burst}=1$ \\
New & dynamic halo & dynamical friction + satellite collisions &
all mergers; $e_{\rm burst}=f(m_1/m_2)$ \\
\hline
\end{tabular}
\end{center}
\label{tab:cm_packages}
\end{table*}
\begin{table*}
\caption{Star Formation/Feedback Packages}
\begin{center}
\begin{tabular}{lccc}
\hline
name & star formation & feedback & reheated gas \\
\hline
Munich & SFR-M & eqn.~\ref{eqn:fb_pl}, 
$\alpha_{\rm rh}=2$ &all stays in halo \\
Durham & SFR-D & eqn.~\ref{eqn:fb_pl}, 
$\alpha_{\rm rh}=5.5$ & all ejected from halo \\
Santa Cruz (fiducial) & SFR-M & disk/halo & 
disk/halo \\
Santa Cruz (high fb) & SFR-M & disk/halo & 
disk/halo \\
Santa Cruz (C) & SFR-C & disk/halo & 
disk/halo \\
\hline
\end{tabular}
\end{center}
\label{tab:sffb_packages}
\end{table*}
\subsection{Dust Absorption}
\label{sec:models:dust}
Absorption of galactic light by dust in the interstellar medium causes
galaxies to appear fainter and redder in the ultraviolet to visible
part of the spectrum. In this paper, we have adopted a simple model of
dust extinction based on the empirical results of \citeN{wh}. These
authors give an expression for the B-band, face-on extinction optical
depth of a galaxy as a function of its blue luminosity:
\begin{equation}
\tau_{B} = \tau_{B,*}\left(\frac{L_{B, i}}{L_{B, *}}\right)^\beta\, ,
\end{equation}
where $L_{B, i}$ is the intrinsic (unextinguished) blue luminoisity,
and we use $\tau_{B, *}=0.8$, $L_{B, *}=6\times 10^9 L_{\odot}$, and
$\beta=0.5$, as found by \citeN{wh}. We then relate the B-band optical
depth to other bands using a standard Galactic extinction curve
\cite{cardelli:89}. The extinction in magnitudes is then related to
the inclination of the galaxy using a standard ``slab'' model (a thin
disk with stars and dust uniformly mixed together):
\begin{equation}
A_{\lambda}=-2.5 \log \left (\frac{1-e^{-\tau_{\lambda} \sec \theta}}
            {\tau_{\lambda} \sec \theta} \right )
\end{equation}
where $\theta$ is the angle of inclination to the line of
sight\cite{guiderdoni:87,tully-fouque:85}. We assign a random
inclination to each model galaxy. The extinction correction is only
applied to the disk component of the model galaxies (i.e. we assume
that the bulge component is not affected by dust).

\subsection{Model Packages}
\label{sec:models:packages}
There are many possible permutations of the different ingredients that
we have introduced above. In the interest of practicality, we have
chosen several ``packages'' of ingredients to explore in this
paper. In the relevant sections, we comment on which elements of the
package are important in determining various quantities.

In order to understand the effects of the new ingredients that we have
introduced pertaining to cooling, galaxy merging, and starbursts, we
introduce two ``cooling/merging'' packages (see
Table~\ref{tab:cm_packages}). The ingredients of the ``Classic''
package were chosen to be similar to the published Munich and Durham
models
\footnote{Except that the cooling model used in the published Durham
models is more similar to our ``dynamic halo'' model (see
Section~\ref{sec:models:cooling}).}. In this paper, we always apply
them within the SCDM cosmology, which was used for much of the
previous work. The ingredients of the ``New'' cooling/merging package
reflect additions or modifications to our models. We apply the ``New''
models within more realistic (or anyway more fashionable)
cosmologies. The new ingredients are described above, namely, the
dynamic-halo cooling model (Section~\ref{sec:models:cooling}),
satellite collisions (Section~\ref{sec:models:mergers}), and more
detailed modelling of starbursts in galaxy-galaxy mergers. In the
``Classic'' models, starbursts occur only in major mergers and with
efficiency $e_{\rm burst}=1$. In the ``New'' models, starbursts occur
in all mergers and $e_{\rm burst}$ is a function of the mass ratio of
the merging galaxies. The details of the starburst modelling are of
minor importance for galaxy properties at $z\sim0$, and will be dealt
with in detail in a companion paper \cite{spf}.

We also wish to understand the effects of different choices of star
formation and supernova feedback recipes, and introduce several
``sf/fb'' packages. We have chosen the ingredients of the first two
packages to be similar to the choices made by the Munich and Durham
groups with respect to the star formation and supernova feedback,
including the fate of reheated gas (see
Section~\ref{sec:models:feedback}). It should be kept in mind,
however, that although we will refer to these as the ``Munich'' and
``Durham'' \emph{packages}, our models differ from those of these
other groups in many respects and we are not trying to reproduce their
results in detail. On the contrary, we wish to isolate the effects of
the way that star formation and supernova feedback are modelled. For
example, as we will discuss in Section~\ref{sec:models:params}, the
published models of the Munich and Durham groups are normalized such
that a galaxy of a given circular velocity is considerably fainter
than in our models. Here we will always normalize all of the packages
in the same way, as described in Section~\ref{sec:models:params}. We
will refer to the results that we obtain from our code, normalized as
described in this paper, as the Munich and Durham ``packages''. When
we wish to refer to the results obtained by the Munich and Durham
groups using their codes, normalized in their own ways, we shall refer
to the ``actual'' or ``published'' Munich or Durham models.

The third package, which we refer to as the ``Santa Cruz (fiducial)''
package, is a hybrid of the Munich-style star formation law (SFR-M)
and the disk-halo feedback model described in
Section~\ref{sec:models:sf}. ``Santa Cruz (high fb)'' is the
same as Santa Cruz (fiducial) except that the supernova feedback
parameter is turned up by a factor of five. The ``Santa Cruz (C)''
package assumes that the star formation efficiency is constant at all
redshifts and in galaxies of all sizes (SFR-C). Note that this is
equivalent to the star formation law suggested for use with milder
supernova feedback ($f_{\rm v}=0.01$) by CAFNZ94. Again, we combine
this with the disk-halo feedback model.

\section{Setting the Galaxy Formation Parameters}
\label{sec:models:params}
We have introduced a number of parameters. ``Free'' parameters are
adjusted for each choice of cosmology and package of astrophysical
recipes, according to criteria that we shall describe. The ``fixed''
parameters keep the same values for all of the models. In this
section, we summarize the parameters and the procedure that we use to
set them.
\begin{figure}
\centerline{\psfig{file=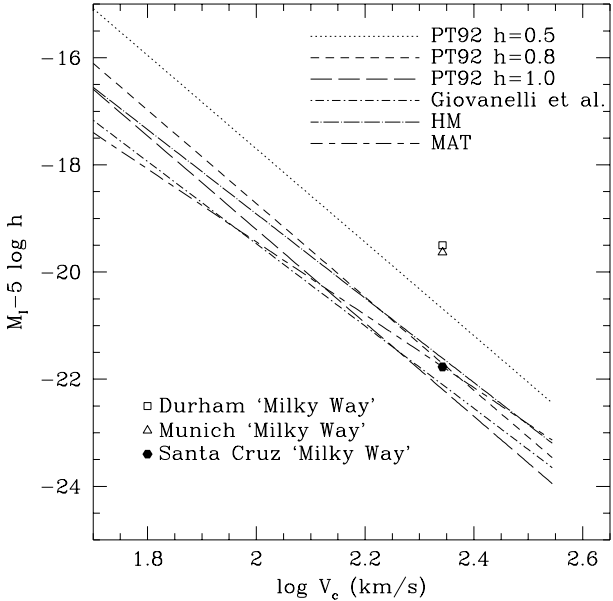,height=8.5truecm,width=8.5truecm}}
\caption{Fits to the observed I-band Tully-Fisher relation from
several samples. PT92 is the \protect\cite{pierce-tully:92} relation
for nearby galaxies, scaled assuming various values of the Hubble
parameter. Giovanelli et al. is from the \protect\citeN{giovanelli:97}
sample of cluster spirals. HM and MAT are the Han-Mould and Mathewson
et al. samples discussed in Willick et
al. \protect\citeyear{willick:I,willick:II}. The magnitude of a
``Milky Way'' galaxy ($V_c=220\, \kms$) in the Durham models, from
Fig.~11 of CAFNZ94 is indicated by the open square. The approximate
magnitude of a ``Milky Way'' galaxy in the Munich models is indicated
by the open triangle. The magnitude of the ``Milky Way'' in our models
is indicated by the filled hexagon. }
\label{fig:tf_obs}
\end{figure}
\subsection{Fixed parameters}
The physical meaning and values of the fixed parameters, as well as
the section in which they are discussed in detail, are summarized as
follows:
\begin{itemize}
\item $\alpha_{*}=-1.5$ (\ref{sec:models:sf}): 
power used in power-law (Durham) star formation law
\item $\alpha_{rh}$ (\ref{sec:models:feedback:previous}):
 power used in supernova reheating power-law ($\alpha_{rh}=2$ 
for Munich, $\alpha_{rh}=5.5$ for Durham) 
\item $f_{\rm rh, disk}=2$ (\ref{sec:models:feedback:disk-halo}) 
: fudge factor used in disk-halo feedback model
\item $f_{\rm reheat}=0.5$ (\ref{sec:models:cooling:mergers}): 
In the dynamic halo cooling model, the cooling time $\tau_{\rm cool}$ is reset
after a halo merger event if the largest progenitor of the current
halo is less than a fraction $f_{\rm reheat}$ of its mass.
\item $Z_{\rm hot}$ (\ref{sec:models:cooling}):
the assumed metallicity of the hot gas, used in the cooling function. 
\end{itemize}

\subsection{Free Parameters}
The free parameters and the sections in which they were introduced are:
\begin{itemize}
\item $\tau_{*}^0$ (\ref{sec:models:sf}): the star formation timescale
\item $\epsilon^0_{\rm SN}$ (\ref{sec:models:feedback}):
 supernova reheating efficiency 
\item $y$ (\ref{sec:models:chemev}) : 
chemical evolution yield (mass of metals produced per 
unit mass of stars)
\item $f^{*}_{\rm lum}$ (\ref{sec:models:popsynth}):
 the fraction of the total stellar mass in luminous stars
\item $f_{\rm mrg}$ (\ref{sec:models:mergers:df}): 
the initial distance of satellite halos from the
central galaxy after a halo merger, in units of the (post-merger)
virial radius.
\item $f_{\rm bulge}$ ({sec:models:morph}): 
the mass ratio that divides major mergers 
	from minor mergers; determines whether a bulge component is
	formed
\end{itemize}

To set the values of the free parameters, we define a fiducial
``reference galaxy'', which is the central galaxy in a halo with a
circular velocity of \refgal. We set the most important free
parameters by requiring the properties of this reference galaxy to
agree with observations for an average galaxy with this circular
velocity. As an important constraint, we would like to require our
reference galaxy to have an average I magnitude given by the
Tully-Fisher relationship (we normalize in I rather than B because it
is less sensitive to recent starbursts and the effects of dust). But
first, we discuss a subtlety in the process of comparing the models
with this observation, which has led to some confusion in the past.

\subsubsection{Tully-Fisher Normalization}
If we use a local sample, such as that of \citeN{pierce-tully:92}, the
relation between absolute magnitude and line-width has been determined
by measuring a distance to each galaxy using various standard methods
(e.g. Cepheids, RR Lyrae, planetary nebulae). This relation therefore
intrinsically contains an effective Hubble parameter. The
\citeN{pierce-tully:92} sample, when used to derive distances to the
Ursa Major and Virgo clusters, implies $H_0 = 85 \pm 10\, \kmsmpc$
\cite{pierce-tully:88}. Our models are set within predetermined
cosmologies with various values of the Hubble parameter ($H_0 =$ 50,
60, or 70$\: \kmsmpc$). To compare this data to the different
cosmologies used in our models, one approach is to simply scale the
observed absolute magnitudes:
\begin{equation}
\label{eqn:magscale}
M_{\rm model} = M_{\rm obs} + 5\log(h_{\rm model}/h_{\rm obs})
\end{equation}
This is effectively what CAFNZ94 say they have done in their Fig.~11,
assuming $h_{\rm obs}=1.0$ (although it looks more as though they used
$h=0.80$). They then interpret their Fig.~11 as indicating that their
models are discrepant with the observed Tully-Fisher relation because
their galaxies are $\sim 1.8$ magnitudes too faint at a given circular
velocity. In contrast, KWG93 claim good agreement with the
Tully-Fisher relation, and show this in their Fig.~7. This leaves one
with the impression that a Durham galaxy would be about 1.8 magnitudes
fainter than a Munich galaxy with the same circular velocity.

It is difficult to make a direct comparison from the published papers
because KWG93 plot the TF relation in the B band, and in terms of
luminosity, whereas CAFNZ94 use the I band, and plot $M_I-5\log
h$. However, we can easily check what would happen if we scaled the
B-band relation used by KWG93 in the same way. If we assume $M_{\rm
B,obs}=-20.7$ for a \refgal\ galaxy, from the \citeN{pierce-tully:92}
relation, and take $h_{\rm obs}=1.0$ and $h_{\rm model}=0.5$, this
would imply $M_{\rm B, model}=-22.2$. This is 1.2 to 2.2 magnitudes
brighter than the ``Milky Way'' normalization ($M_B \sim -20$ to -21),
used in the published Munich models. What this means is that the
apparent good agreement with the TFR seen in Fig.~7 of KWG93 is
because they assumed $h_{\rm obs} = h_{\rm model} = 0.5$.

\begin{table}
\begin{center}
\caption{Galaxy formation parameters for the ``Classic'' models (SCDM)}
\begin{tabular}{|lccccc|}
\hline
model & $\tau^{0}_{*}$ & $\epsilon^{0}_{\rm SN}$ &$y$ & $f^{*}_{\rm
lum}$ & $f_{\rm bar}$ \\
\hline
Munich & 100 & 0.2 & 1.3 & 1.0 & 0.1 \\
Durham & 4.0 & N/A & 1.8 & 1.0 & 0.125 \\
Santa Cruz (fiducial) & 100 & 0.1 & 1.8 & 1.0 & 0.125 \\
Santa Cruz (high fb) & 100 & 0.5 & 4.0 & 1.0 & 0.125 \\
Santa Cruz (C) & 8.0 & 0.125 & 1.8 & 1.0 & 0.125 \\
\hline
\end{tabular}
\label{tab:params_sf}
\end{center}
\end{table}
One lesson in all of this is that trying to normalize the models with
the local Tully-Fisher data is problematic because it is so sensitive
to the Hubble parameter. A more robust approach is to use the
\emph{velocity-based} zero-point from the compilation of several more
distant TF samples from Willick et al. \citeyear{willick:I,willick:II}
and \citeN{giovanelli:97}. This effectively gives us a relation
between $M_{I}-5\log h$ and line-width. The Hubble parameter is
explicitly scaled out, so we can apply the normalization fairly across
models with different values of $H_0$. We show the fits to these
observed relations along with the I magnitude of a ``Milky Way''
galaxy for the published Munich and Durham models and for our fiducial
models in Fig.~\ref{fig:tf_obs}. The magnitude of the Durham ``Milky
Way'' is taken from Fig.~11 of CAFNZ94. The Munich ``Milky Way''
is the I-band magnitude that we get in the version of our code in
which we try to reproduce all the assumptions of the Munich models,
and set the free parameters to get the B magnitude they quote ($M_B
\sim -20$). It is approximately the same as in the actual Munich
models (G. Kauffmann, private communication). We can now see that if
placed side by side, the Munich and Durham model galaxies have almost
the same magnitudes at \refgal, and that both are about 2 magnitudes
fainter than the observed I-band TFR, independent of assumptions
about the Hubble parameter. This can be reconciled with the Munich
group's assertion that they reproduce the observed B-band TFR by two
factors. One is the scaling with $H_0$ that has already been
discussed. The second factor is that the model galaxies are too blue
in B-I. We also see that the local
\citeN{pierce-tully:92} relation only agrees with the results of more
distant samples if a relatively high value of the Hubble parameter ($h
\sim 0.85$) is assumed. This is further evidence that some sort of
rescaling is necessary if the local relation is to be used in
conjunction with theoretical models.

Therefore the results of the published Munich and Durham models are
actually more consistent than it appeared. The reason their
``Milky Way'' is so much fainter is easy to understand. The value of
the parameter that we refer to as $f^{*}_{\rm lum}$ is 0.5 in the
Munich models and 0.37 in the Durham models. This corresponds to the
assumption that 50\% or 63\% of the stellar mass is in the form of
non-luminous brown dwarfs or planets. This is an unrealistically large
contribution from non-luminous stars according to most theories of
star formation, and results in stellar mass-to-light ratios about a
factor of 2-3 higher than the observed values
\cite{gilmore:97}. Taking $f^{*}_{\rm lum}=1$ results in more
reasonable stellar mass-to-light ratios, and brings the reference
galaxy into better agreement with the TFR.

\begin{table}
\begin{center}
\caption{Galaxy formation parameters for the ``New'' 
(Santa Cruz fiducial) models}
\begin{tabular}{|lccccc|}
\hline
model & $\tau^{0}_{*}$ & $\epsilon^{0}_{\rm SN}$ &$y$ & $f^{*}_{\rm
lum}$ & $f_{\rm bar}$\\
\hline
SCDM & 100 & 0.1 & 1.7 & 1.0 & 0.125 \\
\taucdm & 100 & 0.05 & 1.8 & 1.0 & 0.11 \\
$\Lambda$CDM.5 & 100 & 0.125 & 3.5 & 1.0 & 0.11\\ 
OCDM.5 & 100 & 0.125 & 3.5 & 1.0 & 0.11\\ 
$\Lambda$CDM.3 & 50 & 0.125 & 2.2 & 0.80 & 0.13\\
OCDM.3 & 80 & 0.125 & 3.7 & 0.9 & 0.13 \\
\hline
\end{tabular}
\label{tab:params_cosmo}
\end{center}
\end{table}
\subsubsection{Setting the Free Parameters}
We now set our parameters to get the central galaxy in a \refgal\ halo
to have $M_{I}-5\log h \sim -21.6$ to -22.1, which is consistent with
the values predicted from the fitted relations for the three distant
I-band surveys mentioned above. Because the observations have been
corrected for dust extinction, we use the \emph{non} dust-corrected
magnitude of the reference galaxy to normalize the models (actually we
should use magnitudes with face-on dust corrections, but in the I-band
these are quite small). To convert between the measured ${\rm H_{I}}$
line-widths $W_{R}^i$ and the model circular velocities, we assume
$W_{R}^i = 2V_c$. However, it should be kept in mind that this
transformation is not necessarily so straightforward, and this could
change the slope and curvature of the relation especially on the
small-linewidth/faint end. Note that we have also implicitly assumed
that the rotation velocity of the galaxy is the same as that of the
halo, i.e. that the rotation curve is flat all the way out to the
virial radius of the halo. This neglects the effect of the
concentrated baryons in the exponential disk, which will increase the
circular velocity at small radii. Moreover, if the dark matter halo
profiles resemble the form found by \citeN{nfw:96}, for galaxy-sized
halos, the rotation velocity at $\sim2$ disk scale lengths (where the
observed TFR is measured) is 20-30\% higher than that at the virial
radius of the halo. This would mean that our \refgal\ galaxy would
live inside a $V_c \sim 180\, \kms$ halo. Both of these effects would
lead to a
\emph{larger} galaxy circular velocity for a given halo mass, hence to
smaller mass-to-light ratios.

We also require our average reference galaxy to have a cold gas mass
of $m_{\rm cold} \sim 3.2 \times 10^{9}$ to $2.5 \times 10^{10} h^{-2}
\msun$. This is consistent with the average ${\rm H_{I}}$ mass of a galaxy 
with $M_{I}-5\log h
\simeq -21.8$ \cite{deblok:96b}, multiplied
by a factor of two to account for molecular hydrogen
\cite{young:89}. This fixes the two main parameters $\tau_{*}^{0}$ and
$\epsilon^0_{\rm SN}$, although there is some unavoidable degeneracy
(see Section~\ref{sec:varyparam}). The yield $y$ is set by requiring
the stellar metallicity of the reference galaxy to be equal to
solar. Note that the value of $y$ does not affect any other properties
of the galaxies because we have used a fixed hot gas metallicity (see
Section~\ref{sec:models:cooling}).

Following \citeN{bcf:96a}, the parameter $f_{\rm bulge}$ is fixed by
requiring the fraction of morphological types to be approximately
E/S0/S+Irr = 13/20/67 (these ratios were obtained by scaling the
observations of \citeN{loveday:96} to account for unclassifiable
galaxies). Using $f_{\rm bulge}=0.25$ results in roughly these
fractions for all the models investigated here, and this value is used
throughout this paper.

The values of the free parameters used in the models presented in this
paper are given in Table~\ref{tab:params_sf} and
Table~\ref{tab:params_cosmo}. We run many realizations and use the
average values of these quantities in order to fix the values of the
free parameters. 

\begin{figure}
\centerline{\psfig{file=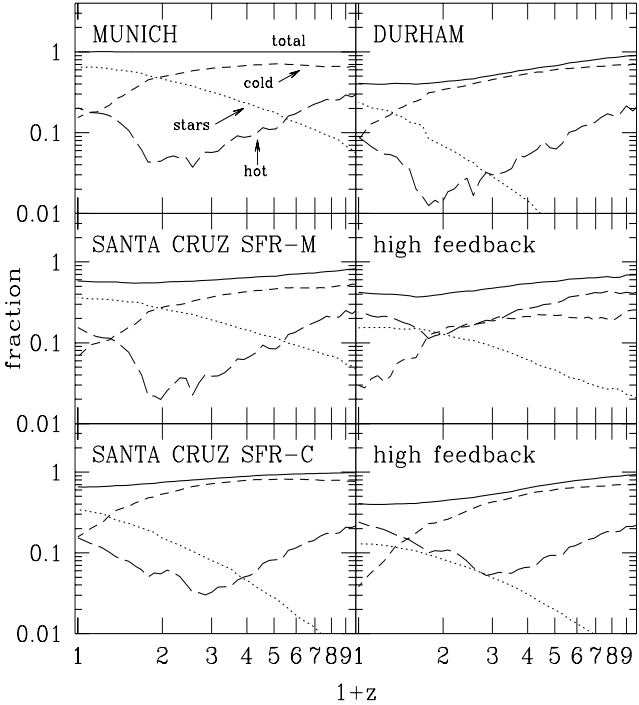,height=10truecm,width=8.9truecm}}
\caption{The history of stars, cold and hot gas within all halos that
will eventually form a ``local group'' ($V_c = 220 \kms$) sized halo
at $z=0$ in the Classic/SCDM models. The solid lines indicate the total
baryon fraction (stars + cold + hot) within the halo, with respect to
the universal value.  Dotted, short, and long dashed lines indicate the
fraction of baryons in the form of stars, cold gas, and hot gas
respectively.}
\label{fig:lghistory_sf}
\end{figure}
\section{The Formation of an $L_*$ Galaxy}
\label{sec:varyparam}
As we have discussed, we normalize our models by requiring certain
properties of a reference galaxy about the size and luminosity of the
Milky Way to agree with observations. In this section, we illustrate
how the formation history of our reference galaxy and its satellite
companions depends on the prescriptions we use for star formation and
supernova feedback (sf/fb), and the values of our free
parameters. This will help in interpreting the results of the next
section, in which we show how global quantities such as the luminosity
function depend on these assumptions.

\begin{figure}
\centerline{\psfig{file=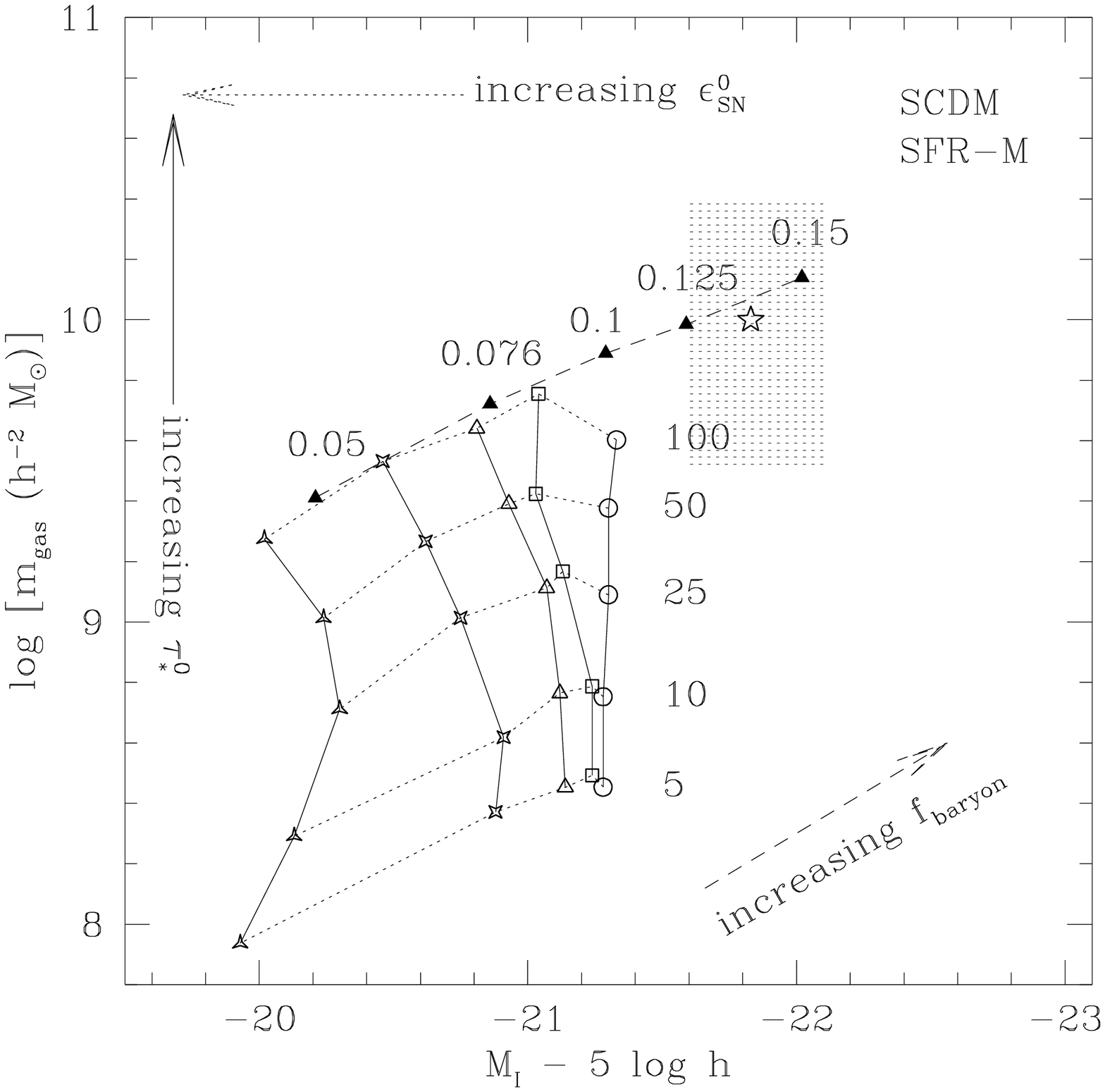,height=8.5truecm,width=8.5truecm}}
\centerline{\psfig{file=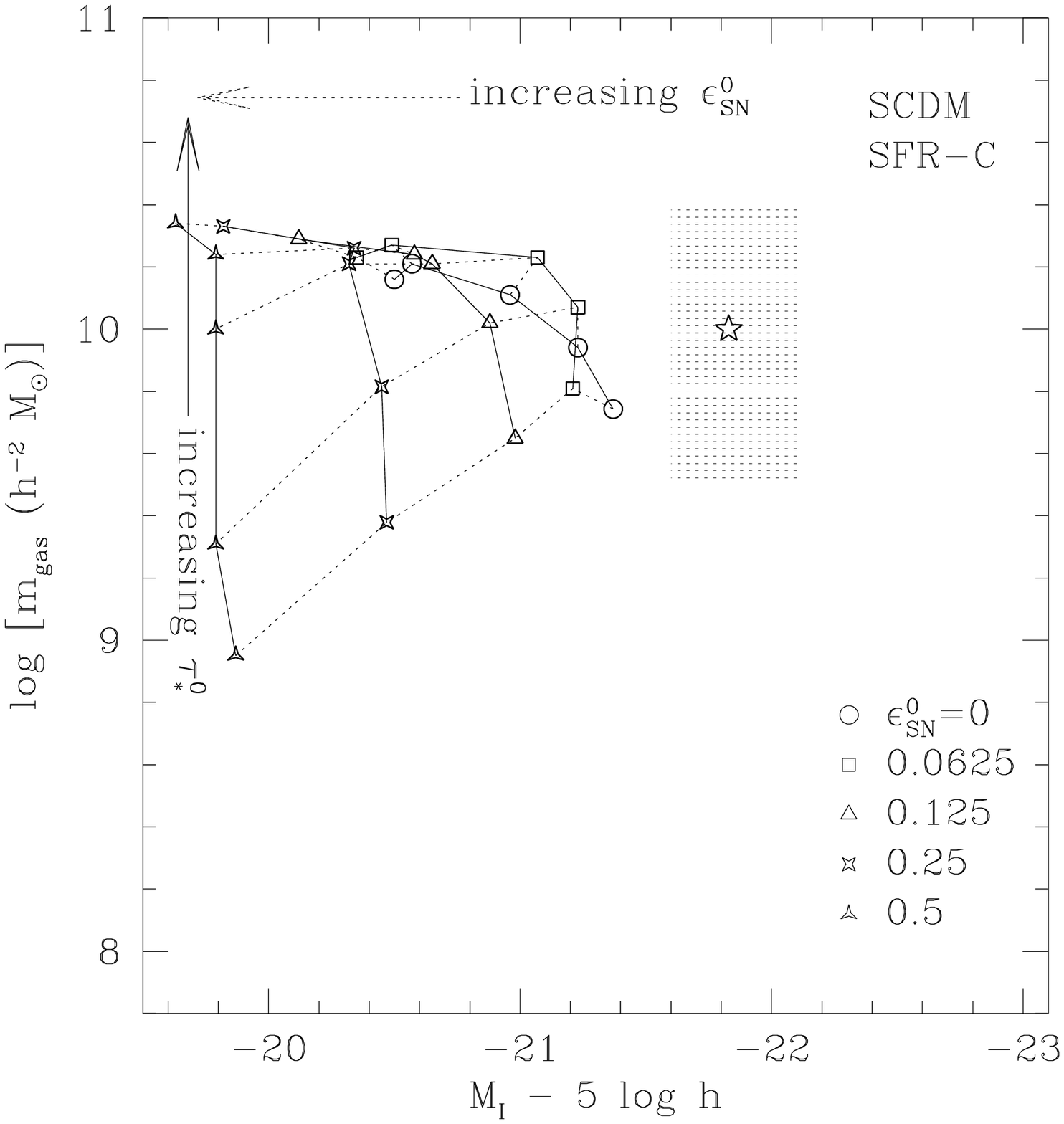,height=8.5truecm,width=8.5truecm}}
\caption{The change in I-band magnitude and cold gas mass of an average
reference galaxy in the Classic/Santa Cruz models (top panel: fiducial;
bottom panel: SFR-C), SCDM cosmology, as the free parameters are tuned. As
$\tau_{*}^0$ is increased, one moves upwards along the solid lines
connecting symbols of the same shape. As $\epsilon_{\rm SN}^0$ is
increased, one moves leftwards along the dotted lines and the symbol
shapes change. This ``grid'' is run with fixed $f_{\rm
baryon}=0.076$. The dotted line connecting filled triangles (top
panel) shows the
effects of varying the baryon fraction $f_{\rm baryon}$, with fixed
$\tau_{*}^0=100$ and $\epsilon_{\rm SN}^0=0.25$.}
\label{fig:parbox_sffb}
\end{figure}
\begin{figure}
\centerline{\psfig{file=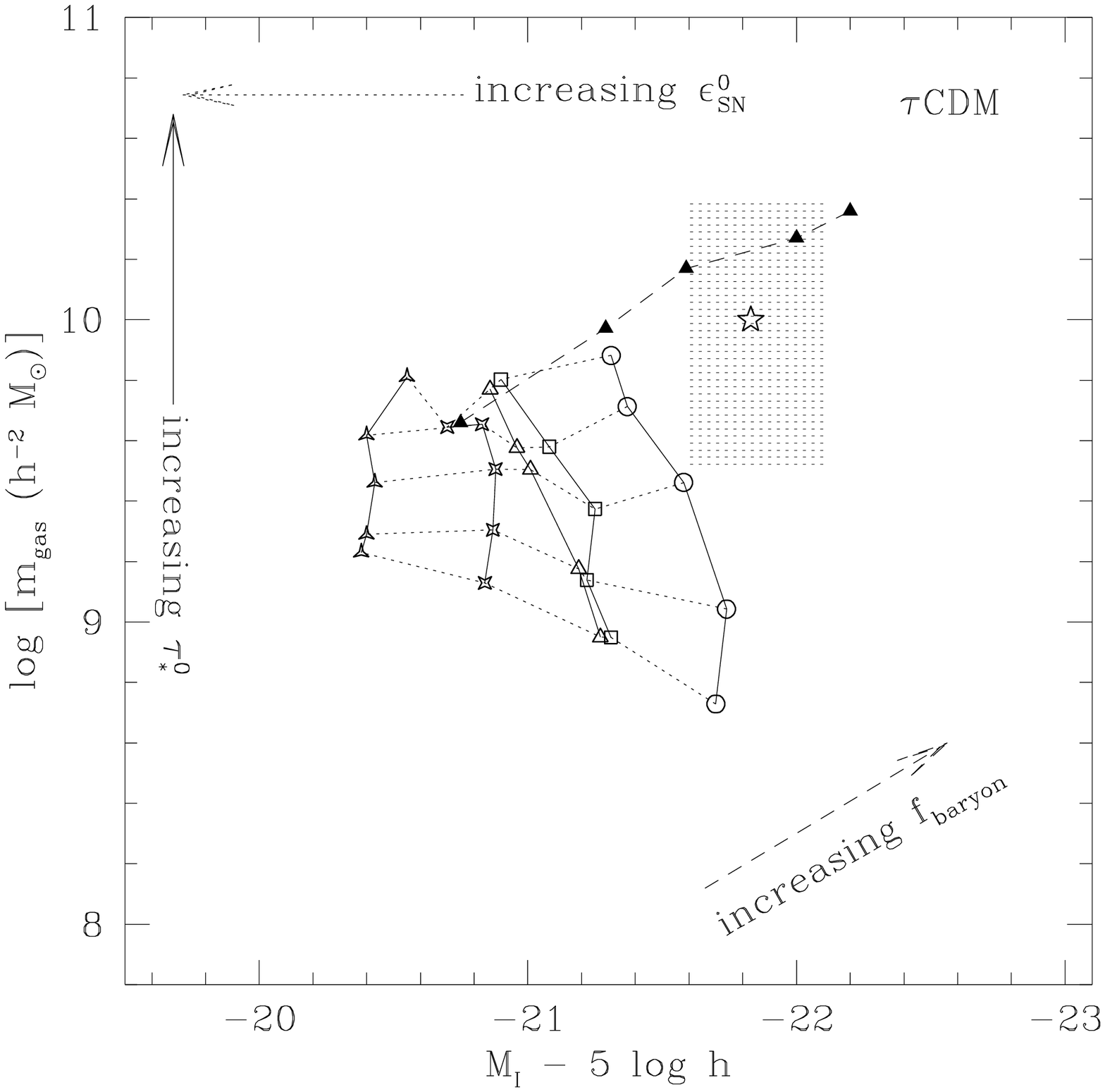,height=8.5truecm,width=8.5truecm}}
\centerline{\psfig{file=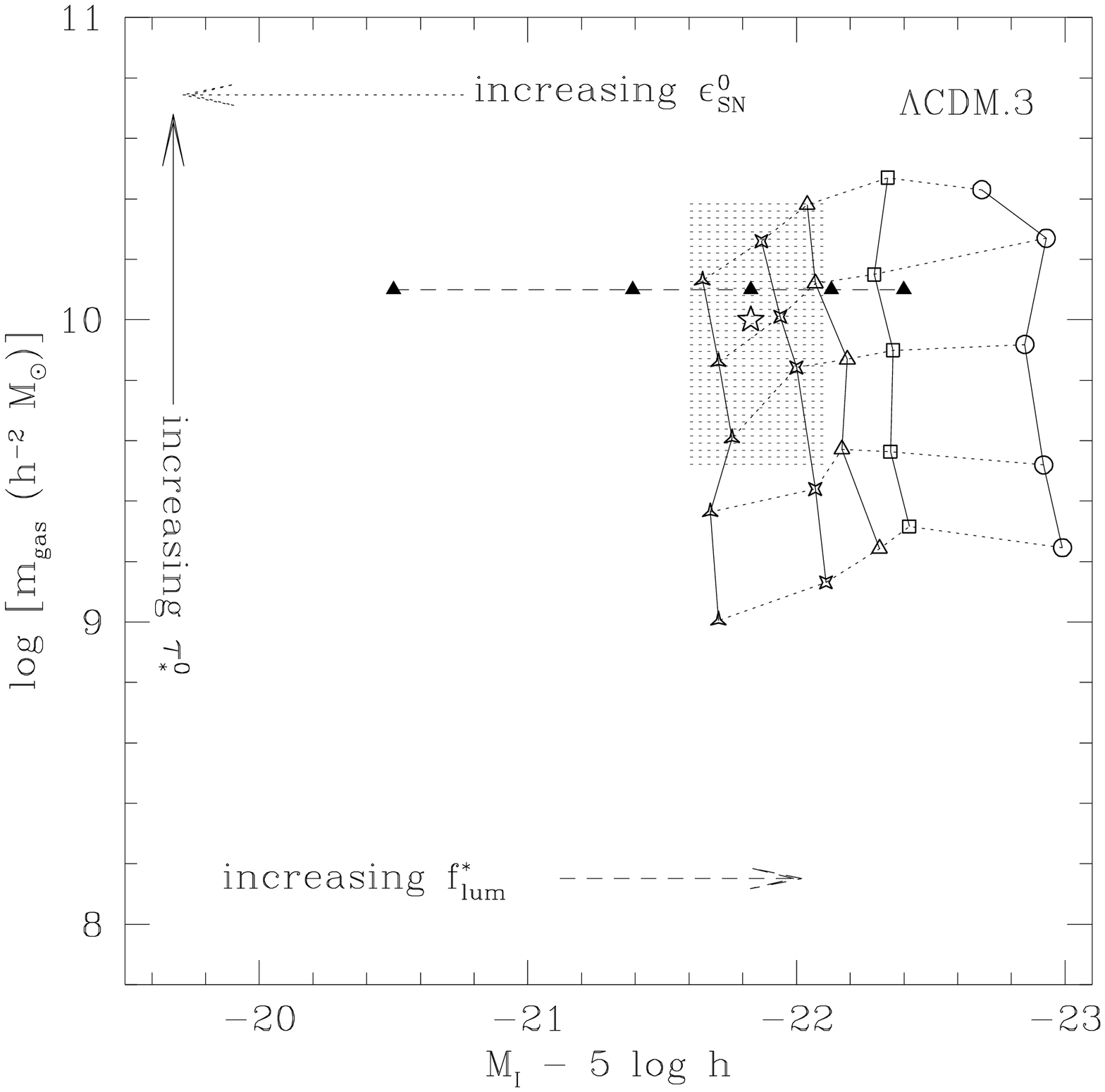,height=8.5truecm,width=8.5truecm}}
\caption{The change in I-band magnitude and cold gas mass of an average
reference galaxy in the New/Santa Cruz (fiducial) models, \taucdm\
and \lcdm3 cosmologies, for different values of the free
parameters. Open symbols connected by dotted and solid lines represent
the same values of $\tau_{*}^0$ and $\epsilon_{\rm SN}^0$ as in
Fig.~\protect\ref{fig:parbox_sffb}, with $f_{\rm baryon}=0.076, 0.129$
(\taucdm, \lcdm3) and $f_{\rm lum}^{*}=1$. Filled triangles in the top
panel represent varying the baryon fraction as in
Fig.~\protect\ref{fig:parbox_sffb} (top panel). In the bottom panel,
the filled triangles represent varying values of $f_{\rm lum}^{*}=0.2,
0.4, 0.6, 0.8, 1.0$ (left to right).}
\label{fig:parbox_cosmo}
\end{figure}
\begin{figure}
\centerline{\psfig{file=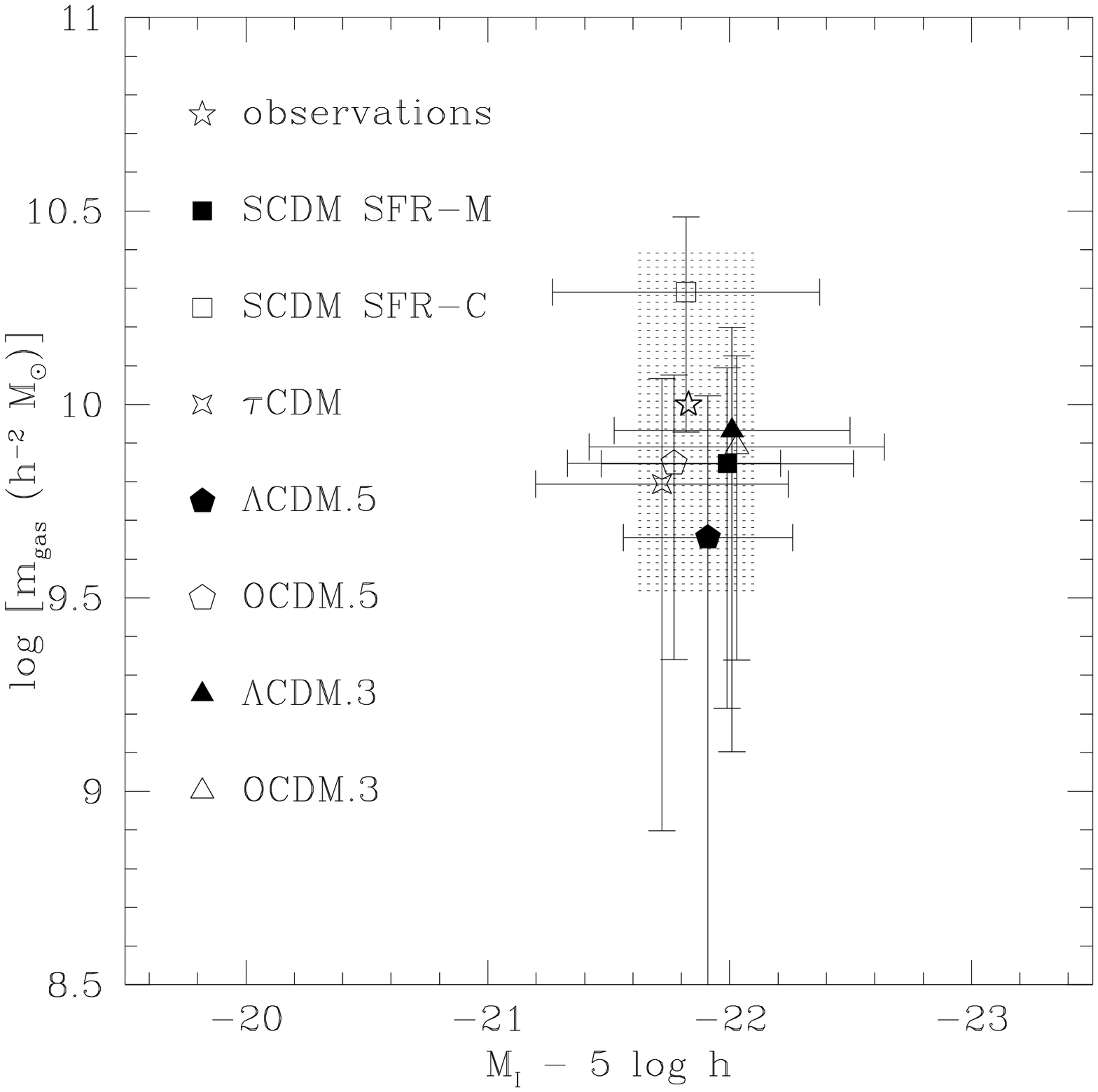,height=8.5truecm,width=8.5truecm}}
\caption{Average I-band magnitude and cold gas mass of a fiducial 
reference galaxy in the Santa Cruz models. Error bars indicate the
standard deviation in these quantities over many realizations.}
\label{fig:fidgal}
\end{figure}
Table~\ref{tab:params_sf} shows the fiducial values of the free
parameters used for each of the sf/fb packages introduced in
Section~\ref{sec:models:packages}. The ``Classic'' cooling/merging package
and the SCDM cosmology are used for all of these models, and the
dynamical friction parameter is set to $f_{\rm
mrg}=1$. Table~\ref{tab:params_cosmo} shows the parameters used for
the models with the ``New'' cooling/merging package and the Santa Cruz
(fiducial) sf/fb package. For these models, we set $f_{\rm mrg}=0.5$,
which is in better agreement with the results of high resolution
N-body simulations (A. Klypin, private communication). It should be
noted that the specific values of these parameters may depend somewhat
on the details of the implementation of our code. Also note that
$\tau^{0}_{*}$ and $\epsilon^{0}_{\rm SN}$ do not function in
precisely the same way in the different packages because of the
differing functional forms of the recipes, so they cannot always be
compared directly.

Fig.~\ref{fig:lghistory_sf} illustrates the redshift evolution of the
baryonic content (stars, cold gas, and hot gas) of halos that will
eventually form a ``local group'' (\refgal) sized halo at $z=0$. The
dependence on the sf/fb package and the value of the supernova
feedback parameter is shown. Note that star formation occurs much
earlier in the Munich package than in the Durham package models. This
is due to two combined effects. First, as we discussed in
Section~\ref{sec:models:sf}, with SFR-M the star formation efficiency
is higher at high redshift because the typical galaxy dynamical times
are shorter. In SFR-D, star formation is less efficient in objects
with smaller circular velocities. At high redshift the characteristic
circular velocities tend to be smaller, so this leads to less star
formation. Second, the much stronger supernova feedback in the
Durham package models leads to additional suppression of star
formation, especially in small objects.

The bottom four panels break down these ingredients. In the Santa Cruz
(fiducial) package, we use SFR-M. The effect of turning up the
feedback efficiency by a factor of five is shown in the right
panel. Star formation is suppressed, and more so at higher redshift
where objects are smaller, but the effect is not as dramatic as in the
Durham package. The bottom-most panels show the Santa Cruz (C)
package, which assumes constant star formation efficiency (SFR-C). This
package is intermediate between the Durham package and the Santa Cruz
SFR-M (fiducial) package.

Fig.~\ref{fig:parbox_sffb} shows how tuning the free parameters
changes the properties of the reference galaxy in the ``Classic''
Santa Cruz (fiducial and C) models, within the SCDM cosmology. The
figure shows the space of I-band magnitude and cold gas mass, along
with the target area used to normalize the models (shaded
box). Symbols show the location of average reference galaxies within
this space for different values of the free parameters $\tau_{*}^0$,
$\epsilon_{\rm SN}^0$, and $f_{\rm baryon}$. The dependence on the
free parameters takes a different form for different star
formation/feedback recipes.  Generally, increasing the star formation
timescale $\tau_{*}^0$ leads to an increased gas mass, and to a much
lesser extent, a fainter luminosity. Increasing the supernova
feedback efficiency $\epsilon_{\rm SN}^0$ leads to a fainter
luminosity and, to a lesser extent, smaller gas mass. Increasing
$f_{\rm baryon}$ leads to a larger gas mass and luminosity. The same
exercise is repeated in Fig.~\ref{fig:parbox_cosmo} with the New/Santa
Cruz (fiducial) package, for the \taucdm\ and \lcdm3 cosmologies. Here
we show the effect of varying $f_{\rm lum}^{*}$, which can only make
galaxies fainter as it can only take values less than
one. Figure~\ref{fig:fidgal} shows the location of the average
reference galaxy in this space and the standard deviation of these
quantities over many ensembles, for all the Santa Cruz models, for the
final fiducial values of the free parameters shown in
Tables~\ref{tab:params_sf} and \ref{tab:params_cosmo}.

We would have liked to consider $f_{\rm baryon}$ to be determined
independently, thus eliminating a free parameter. However, it is
apparent from Fig.~\ref{fig:parbox_sffb} that if we take $f_{\rm
baryon}=0.076$, which corresponds to the baryon fraction derived from
observations of deuterium at high redshift, $\Omega_b h^{2} = 0.019$
\cite{tytler:99} for $h=0.5$ and $\Omega_0=1$, the reference galaxy
is too faint and gas poor in the $\Omega_0=1$ cosmologies compared to
our desired normalization. The values of $f_{\rm baryon} \simeq 0.11$
to $0.13$ that we find necessary to obtain our desired normalization
are similar to those typically derived from very different
considerations in groups and clusters \cite{mohr:99}. As emphasized by
\citeN{white:93}, for high values of $\Omega_0\simeq1$ this is
inconsistent with Big Bang Nucleosynthesis
\cite{bbn}, and with the measurement of \citeN{tytler:99}.
This could be interpreted as further evidence from the galaxy side
that $\Omega_0$ is probably less than unity. It is
\emph{curious} that the best agreement occurs for 
$\Omega_0\simeq 0.4-0.5$ and $h=0.6-0.65$, just the values currently
favored by independent considerations. But given the large and
numerous uncertainties in our modelling (particularly cooling,
feedback, star formation efficiency, and the IMF), we do not regard
this as much more than a curiousity, albeit a rather comfortable
one. For example, we have neglected the eventual return of the gas
expelled by supernovae, and the recycled gas from dead stars. If these
were included, we might be able to reduce the value of $f_{\rm
baryon}$ somewhat. For the moment, we formally consider $f_{\rm
baryon}$ and $f_{\rm lum}^{*}$ to be simply free parameters, which are
close enough to their plausible physical values as to not cause too
much concern.

\begin{figure}
\centerline{\psfig{file=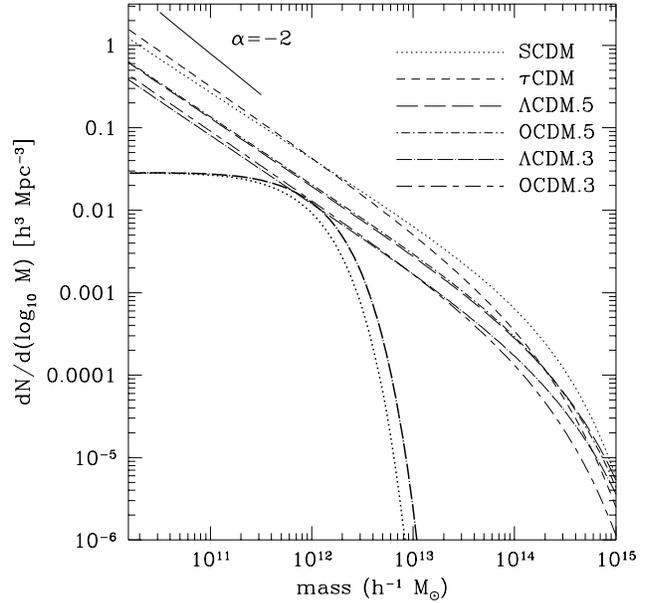,height=8.5truecm,width=8.5truecm}}
\caption{The mass function of dark matter halos predicted by the
standard Press-Schechter model for various CDM cosmologies (light
broken lines). The bold lines show the mass function of galactic
halos, estimated from the observed APM luminosity function as
described in the text, for an SCDM or \taucdm\ cosmology (dotted), and
for the
\lcdm3 cosmology (long dashed-dotted line); 
other cosmologies lie between these two cases). The short solid line
shows a power-law with slope $\alpha=-2$.}
\label{fig:massfunc}
\end{figure}
\section{Comparison with Local Observations}
\label{sec:results}
In this section, we investigate the predictions of our models for a
number of important galaxy properties. We have several goals: we
compare the results of our models with previously published work,
explore the importance of the choice of sf/fb recipe and the values of
certain free parameters, and compare with observational results.

\begin{figure}
\centerline{\psfig{file=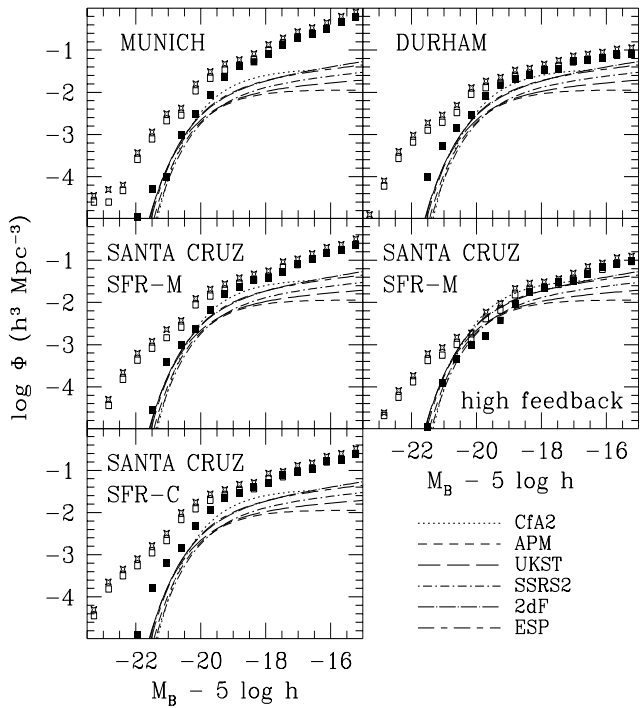,height=10truecm,width=8.9truecm}}
\caption{The B-band luminosity function of galaxies for the Classic/SCDM
models. Crosses and open squares show the models with the original
Press-Schechter weighting and the improved Press-Schechter weighting
of \protect\citeN{sheth-tormen}, both without dust extinction. Solid
squares show the models with inclusion of the empirical dust models
and the improved Press-Schechter model. Dashed lines indicate the fits
to the observed luminosity function from several redshift surveys as
indicated in the key (references given in the text). }
\label{fig:lfb_sf}
\end{figure}
\begin{figure}
\centerline{\psfig{file=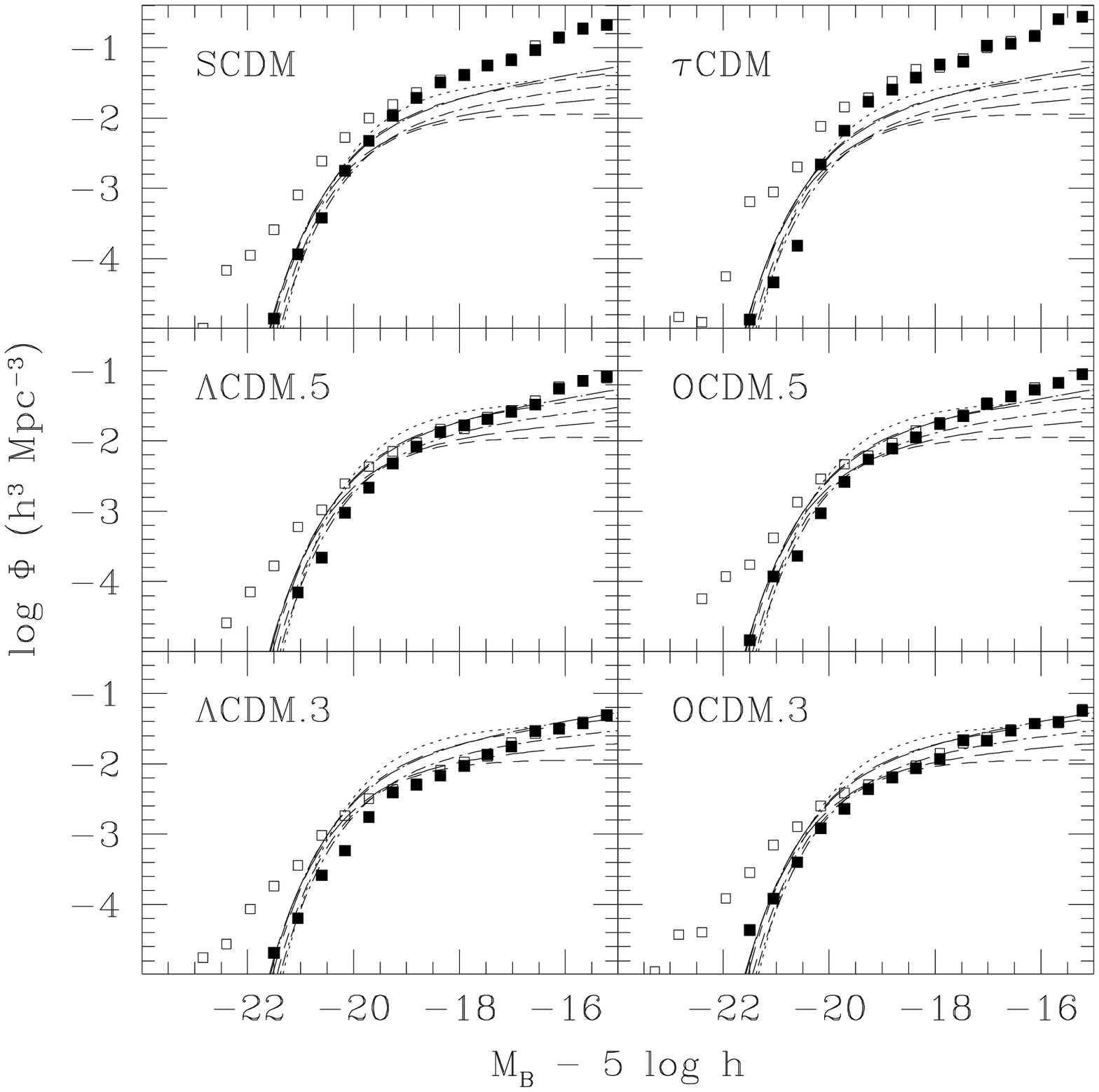,height=10truecm,width=8.9truecm}}
\caption{The B-band luminosity function of galaxies for the New/Santa Cruz 
(fiducial) models. Key as in Fig.~\protect\ref{fig:lfb_sf}. }
\label{fig:lfb_cosmo}
\end{figure}
\begin{figure}
\centerline{\psfig{file=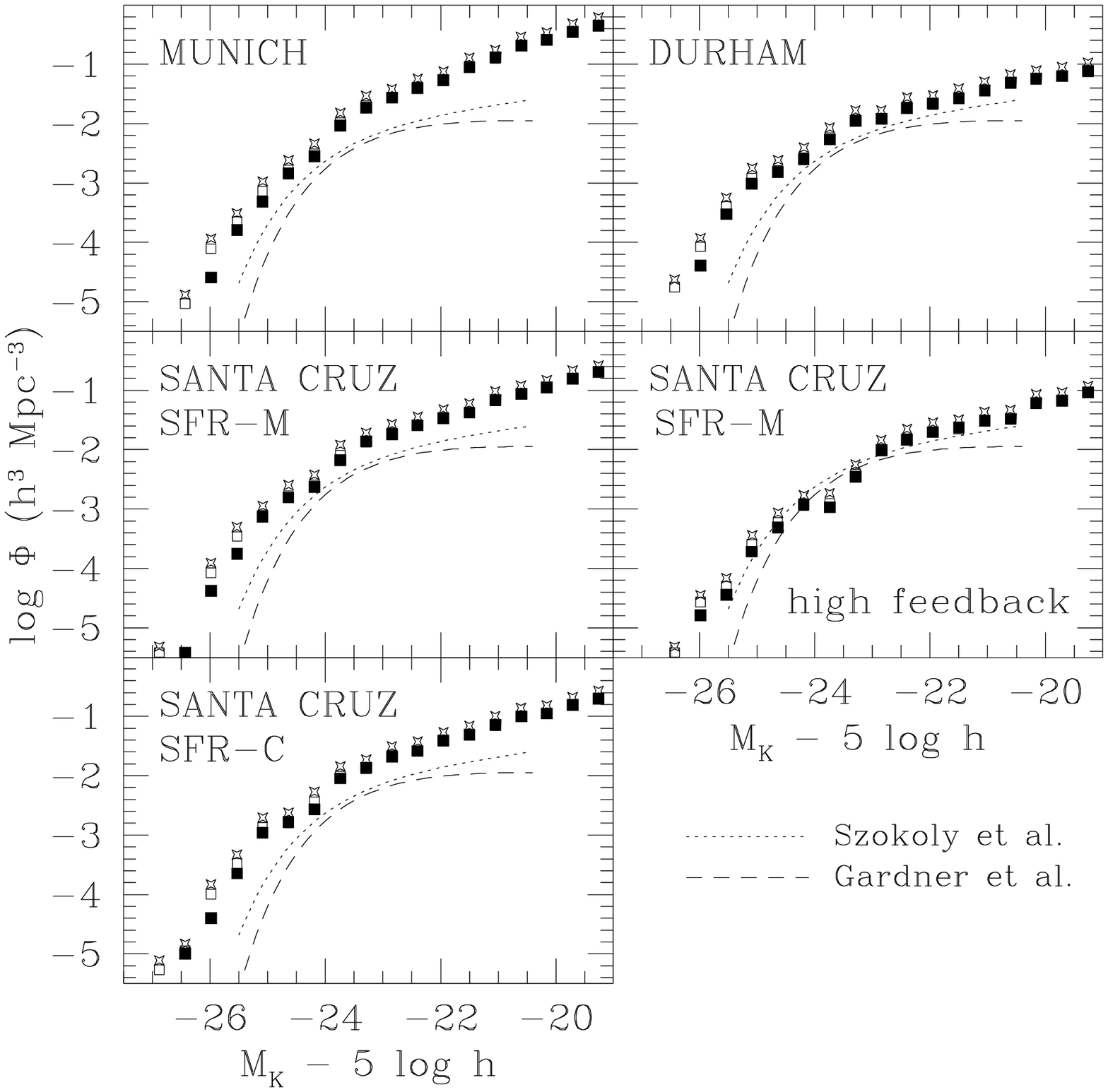,height=10truecm,width=8.9truecm}}
\caption{The K-band luminosity function of galaxies for the Classic/SCDM
models. Crosses and open squares show the models with the original
Press-Schechter weighting and the improved Press-Schechter weighting
of \protect\citeN{sheth-tormen}, both without dust extinction. Solid
squares show the models with inclusion of the empirical dust models
and the improved Press-Schechter model. Dashed lines indicate the fits
to the observed luminosity function as indicated in the key
(references given in the text). }
\label{fig:lfk_sf}
\end{figure}
\begin{figure}
\centerline{\psfig{file=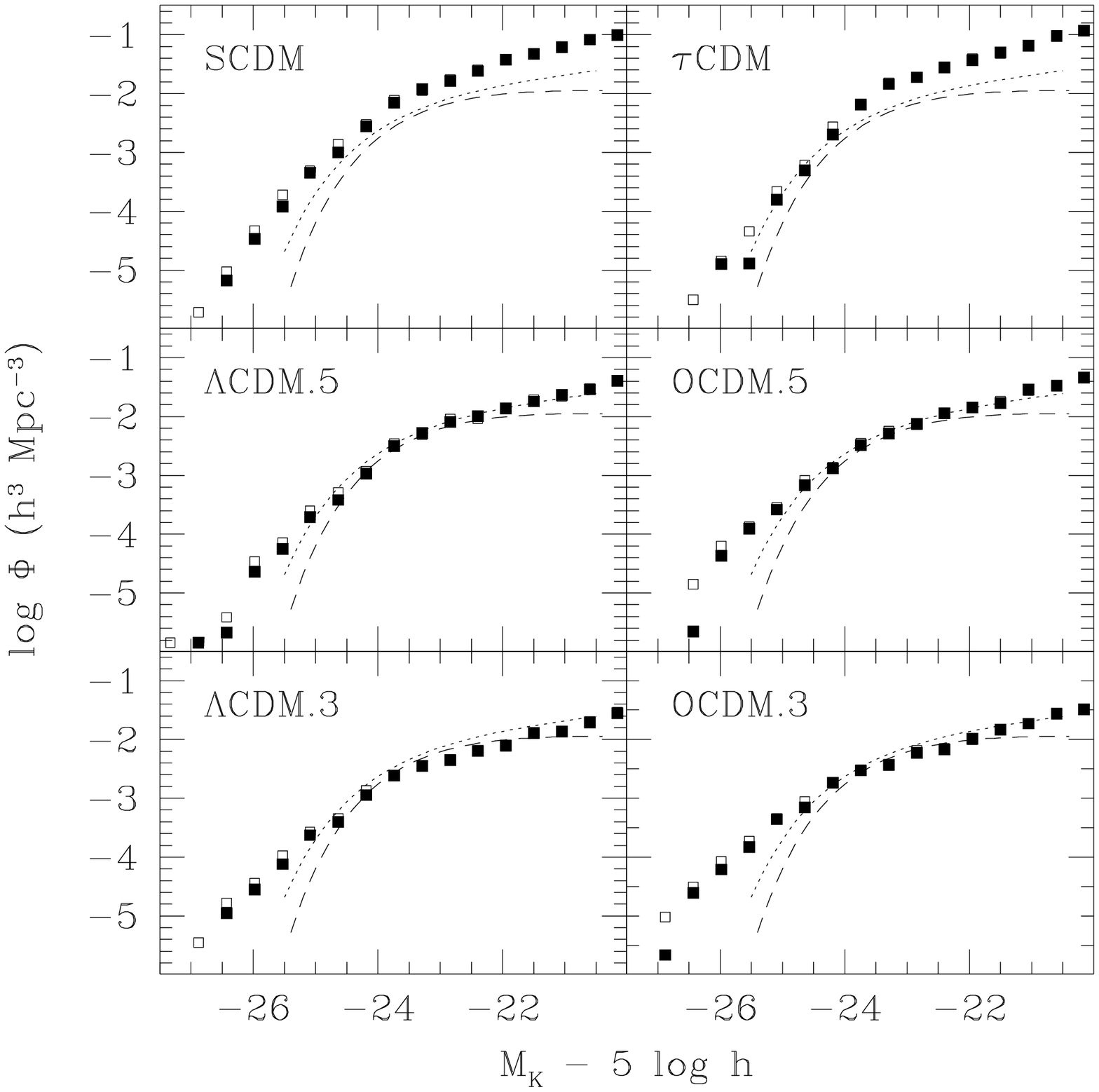,height=10truecm,width=8.9truecm}}
\caption{The K-band luminosity function of galaxies for the New/Santa Cruz 
(fiducial) models for different cosmologies. Key as in 
Fig.~\protect\ref{fig:lfk_sf}. }
\label{fig:lfk_cosmo}
\end{figure}
The local number density of galaxies as a function of their luminosity
is clearly a key prediction of any successful model of galaxy
formation. The halo mass function predicted by any of the currently
popular CDM-based models has a very different shape from the
characteristic Schechter form of observed luminosity functions. At
masses less than $10^{13}$ \hmsun, the CDM mass function is a power
law with a slope $\alpha \sim -2$, much steeper than the faint-end
slope of the observed field galaxy luminosity function $\alpha
\simeq -1.0$ to $-1.5$. The exponential cut-off occurs at
$\sim 10^{14}$ \hmsun, much larger than the expected halo mass
corresponding to an $L_{*}$ galaxy. In Fig.~\ref{fig:massfunc}, we
show the halo mass function predicted by the standard Press-Schechter
model, along with the mass function of galactic halos estimated in the
following simple way. We find the circular velocity of the dark matter
halo associated with a typical $L_{*}$ galaxy using the Tully-Fisher
relation and assuming the rotation curve is flat. Using $M_{B}^{*}-5
\log h=-19.5$ (cf. \citeNP{loveday}), and the B-band Tully-Fisher
relation
\cite{pierce-tully:92,tullydust:97} scaled to $h_{\rm obs}=0.80$ (see
Section~\ref{sec:models:params}), we find $V_c^*=160$ \kms. We can
then translate this to a mass using the spherical top-hat model (see
Appendix~A). From Fig.~\ref{fig:massvc} we see that this corresponds
to a halo with a mass of about $1.0\times 10^{12} \msun$ to $1.3
\times 10^{12} \msun$, depending on the cosmology. Using this constant
light-to-mass conversion, we translate the observed B-band luminosity
function
\cite{loveday} to the mass function shown in Fig.~\ref{fig:massfunc}. 
Of course this translation is complicated by sub-structure (each of
the halos in the Press-Schechter model may contain multiple galaxies
of various sizes), as well as by the varying mass-to-light ratio of
galaxies of different morphological types and other
complications. However, these effects will introduce changes of order
a factor of a few, and the discrepancy is much larger. The problem may
be summarized as follows. In order to get from any CDM mass function
to the observed luminosity function, it seems that the conversion from
halo mass to galaxy luminosity must be more complicated than what we
have assumed in this simple calculation; in particular, apparently the
mean mass-to-light ratio must decrease as we move away from
$V_{c}^{*}$ in both directions. On the other hand, the constant
mass-to-light model gives us a perfect power-law Tully-Fisher relation
with the correct slope and zero scatter. Any scatter in the
mass-to-light ratio at fixed $V_c$ will introduce scatter in the TFR,
and any systematic variation with $V_c$ will introduce
curvature. Satisfying both constraints simultaneously has proven to be
a challenge.

For example, the first generation of Munich and Durham models
effectively assumed a stellar mass-to-light ratio a factor of 2-3
times larger than the face-value prediction of the Bruzual-Charlot
models (i.e. $f^{*}_{\rm lum} = 0.5$ in the Munich models and 0.63 in
the fiducial Durham models). This pushed the galaxy mass function
(bold curves in Fig.~\ref{fig:massfunc}) to the right, to the point
where the number density roughly agreed at the ``knee''
($L_{*}$). However, it made the galaxies about 2 magnitudes too faint
compared to the observed Tully-Fisher relation. The Durham group
designed their star formation and supernova feedback models in order
to obtain light-to-mass ratios that decreased rapidly with $V_c$. This
flattened the faint end slope of the luminosity function but led to a
pronounced deviation from the observed power-law shape of the
Tully-Fisher relation (cf. Fig.~11 of CAFNZ). In the following two
sub-sections we discuss our results for these two fundamental observed
quantities.

\subsection{The Luminosity Function}
\label{sec:results:lf}
We show the B-band luminosity functions for the Classic SCDM models in
Fig.~\ref{fig:lfb_sf} (the packages are summarized in
Table~\ref{tab:sffb_packages}). The curves show fits to the observed
B-band luminosity functions derived from the CfA \cite{marzke:94},
APM \cite{loveday}, SSRS \cite{dacosta:94}, ESP \cite{zucca}, UKST
\cite{ratcliffe}, and 2dF \cite{twodf} redshift surveys. 
The observational fits have been converted to the Johnson B filter
band used in our models using the conversion $M_{b_J} = M_{Z}-0.45$
for Zwicky magnitudes
\cite{shanks:84} and $M_{B} = M_{b_J} + 0.2$ (CAFNZ94). We show the
effects of using the improved Press-Schechter weighting from the model of
\citeN{sheth-tormen}, and of correcting for dust extinction using the
recipe discussed in Section~\ref{sec:models:dust}. Clearly both of
these effects help to alleviate the tendency of the models to
overpredict the number density of galaxies. The extinction correction
is larger for luminous galaxies (as a direct result of the
Wang-Heckman recipe), but recall that the correction is only applied
to the disk component of our galaxies. Early type galaxies (which are
defined as having large bulge-to-disk ratios) therefore suffer much
smaller corrections. It appears plausible that extinction due to
dust is an important factor in reconciling the discrepancy between the
modelled B-band luminosity function and Tully-Fisher relation. It
should be noted that the observed B-band luminosity function derived
from any of the above redshift surveys is not corrected for the
effects of dust extinction, and Tully-Fisher work always includes a
correction for both internal and Galactic dust extinction. This has
been ignored in the previous theoretical comparisons that we have
discussed. As a point of reference, note that the correction for
internal dust extinction in M31 ranges from 0.27 magnitudes
\cite{pierce-tully:92} to 1.0 magnitude \cite{bernstein} \emph{in the
I-band}. This is to stress that both the corrections and the
uncertainties associated with dust extinction are large. The
corrections are presumably even larger in the B-band, and for more
inclined galaxies.

However, in our models, dust extinction has a negligible effect in
faint galaxies, and very strong feedback (Durham or Santa Cruz with
high feedback) still seems to be necessary to reproduce the observed
faint-end slope within SCDM. The Santa Cruz SCDM models with more
moderate feedback produce a factor of $\sim 3-4$ excess of SMC-sized
galaxies, which is probably difficult to reconcile with observations,
even accounting for sources of incompleteness such as surface
brightness selection effects
\cite{dss:97,dalcanton:97}.

Fig.~\ref{fig:lfb_cosmo} shows the luminosity function of the other
cosmological models, using the New/Santa Cruz (fiducial)
package. The
\taucdm\ model looks quite similar to the SCDM case and similarly shows
an excess of faint galaxies. We have tried several variations of the
\taucdm\ model shown here in an attempt to correct this. One might
think that lowering the normalization $\sigma_8$ would decrease the
overall number density of galaxies. Actually, $\sigma_8$ mainly
controls the location of the exponential cut-off in the mass
function. As we showed in Fig.~\ref{fig:massfunc}, this lies well
above the scale of galactic halos, and so changing $\sigma_8$ within
the bounds allowed by the observed cluster abundance does not
significantly improve our results. We also tried to reduce the number
of faint galaxies by increasing the merging rate (we decreased the
dynamical friction merging timescale to $f_{\rm mrg}=0.1$), but we
find that this does not improve the faint-end significantly and leads
to a severe excess of bright galaxies.

However, the low-$\Omega$ models, particularly the $\Omega_0=0.5$
models, reproduce the overall shape and normalization of the observed
luminosity function remarkably well. Note that at very faint
magnitudes ($M_B-5 \log h \ga -17$) the observed luminosity function
is not well determined, but there is actually a suggestion of the
steepening faintwards of $M_B-5 \log h \ga -17$ that we see in our
$\Omega_0=0.5$ models \cite{zucca,marzke:94,twodf}. The fit on the
bright end could be improved by adjusting the parameters of our dust
recipe, which we have taken at face value from \citeN{wh}. The
$\Omega_0=0.3$ models show a slight
\emph{deficit} of galaxies around $L_{*}$, even without dust extinction, 
but this is within the uncertainties on the normalization of the
observed luminosity function.

\begin{figure}
\centerline{\psfig{file=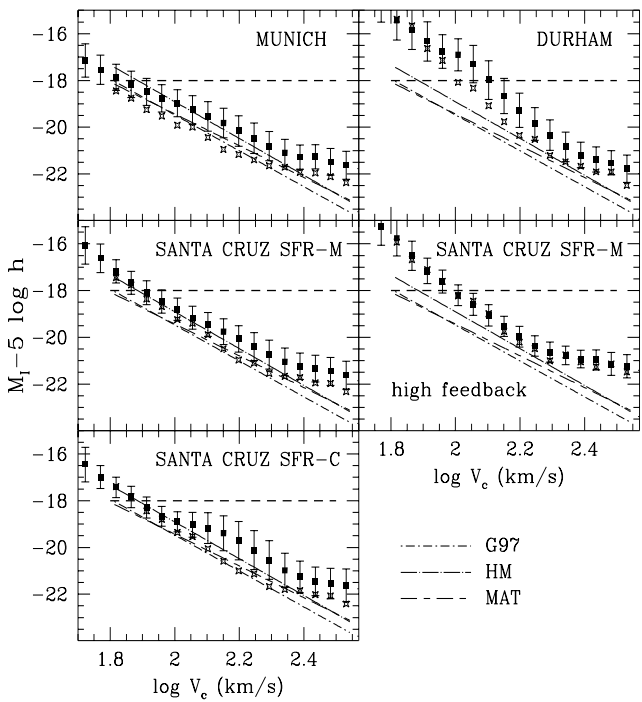,height=10truecm,width=8.9truecm}}
\caption{The Tully-Fisher relation for the Classic/SCDM models. 
Broken lines show fits to the observed I-band Tully-Fisher relation
from several samples (G97 is from
\protect\citeN{giovanelli:97}, HM is the Han-Mould sample, 
and MAT is Mathewson et al. sample from Willick et
al. \protect\citeyear{willick:I,willick:II}). The bold horizontal
dashed line shows the approximate magnitude limit of the
observations. The symbols show the results of the models (crosses show
central galaxies only and filled squares show central and satellite
galaxies), and the error bars indicate 1-$\sigma$ variances over
different merger history realizations. Only model galaxies that
contain cold gas and are identified as spirals are included.}
\label{fig:tf_sf}
\end{figure}
The effects of dust are significantly reduced in longer wavelength
bands such as the near IR, however the observed luminosity function is
not as well determined as it is in optical bands. We compare our
results with two recent determinations of the K-band luminosity
function (we use the K$_{s}$ filter (referred to as $K^\prime$ in the
IRIM manual) downloaded from the KPNO website,
ftp://ftp.noao.edu/kpno/filters, with standard Vega zeropoints). The
Classic/SCDM models are shown in Fig.~\ref{fig:lfk_sf}. The wide-field
K-band survey discussed in \citeN{gardner:97} covers an area of
$\sim4.4$ square degrees, and probably provides the best existing
determination of the bright end of the K-band luminosity function. The
survey discussed in \citeN{szokoly:98} has a smaller area (0.6 square
degrees) but has a fainter limiting magnitude, and thus presumably
provides a more reliable estimate of the faint-end slope.

\begin{figure}
\centerline{\psfig{file=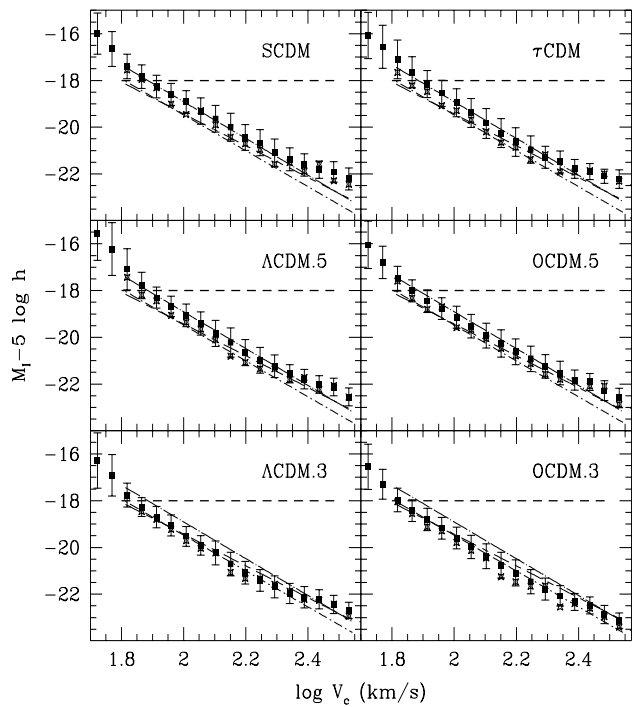,height=10truecm,width=8.9truecm}}
\caption{The Tully-Fisher relation for the New/Santa Cruz (fiducial) 
models. Broken lines show fits to the observed I-band Tully-Fisher
relation from several samples (see Fig.~\ref{fig:tf_sf}). The bold horizontal
dashed line shows the approximate magnitude limit of the
observations. The symbols show the results of the models (crosses show
central galaxies only and filled squares show central and satellite
galaxies), and the error bars indicate 1-$\sigma$ variances over
different merger history realizations. Only model galaxies that
contain cold gas and are identified as spirals are included.}
\label{fig:tf_cosmo}
\end{figure}
All the SCDM models show an overall excess of galaxies of all
luminosities, and the Munich package shows a slightly steeper
faint-end slope than the observations. Both Santa Cruz packages and
the Durham package have a faint-end slope consistent with the
observations of
\citeN{szokoly:98}. The New/Santa Cruz (fiducial) 
models for the other cosmologies are shown in
Fig.~\ref{fig:lfk_cosmo}. The fiducial \taucdm\ models now show a good
match on the bright end but still have an excess on the faint end. The
$\Omega_0=0.5$ models are a near perfect fit over the range of
luminosities probed by the observations, except in the very brightest
bins. They do not cut off as sharply as a pure Schechter function at
brighter luminosities, but the observed luminosity function is not
well determined on the bright end because of small samples and
evolutionary effects. It should be noted that the evolutionary and
k-corrections applied to the data are non-negligible, and are
cosmology dependent. The fits shown here are for a Universe with
$q_0=0.5$, which is inconsistent with our low-$\Omega_0$
cosmologies. A more detailed comparison with the observations is
clearly in order; however, given these uncertainties the level of
agreement shown here is encouraging.

\subsection{The Tully-Fisher Relation}
\label{sec:results:tf}
Recall that we have adjusted the free parameters in our models to
force our reference galaxy to lie on the I band Tully-Fisher
relation derived by Willick et al. \citeyear{willick:I,willick:II} and
\citeN{giovanelli:97}. Fig.~\ref{fig:tf_sf} shows
the fits from the three observational samples mentioned above and the
Tully-Fisher relation we obtain in the Classic/SCDM models. The error
bars indicate the 1-$\sigma$ variance over many Monte-Carlo
realizations. In this plot, we have included only the model galaxies
with more than $10^7 \msun$ of cold gas, and which were classified as
spirals according to their bulge-to-disk ratio as described in
Section~\ref{sec:models:morph}. This is an attempt to select the model
galaxies that most closely correspond to the galaxies in the
observational Tully-Fisher samples we are considering. The Munich and
Santa Cruz packages with moderate feedback produce fairly good
agreement with the slope and scatter of the observed TFR. Note that
central galaxies tend to be brighter than the satellite galaxies. This
is due to our assumption that all new cooling gas is accreted by the
central galaxy, which may not be realistic. We intend to investigate
this using hydro simulations. The Durham package and the Santa Cruz
package with high feedback both show curvature on the faint end due to
the strong supernova feedback. The curvature on the bright end of all
of the Classic models occurs due to the static halo cooling model.

Fig.~\ref{fig:tf_cosmo} shows the TFR for the New/Santa Cruz
(fiducial) models. The results are quite good for all of these
models. There is still a slight curvature on the bright end, but it is
less pronounced, almost absent, in the low-$\Omega$ models. The models
also show a bit of curvature at the very faint end, but this is beyond
the level probed by the observations currently under
consideration. Comparison with samples that probe the TFR to fainter
magnitudes is an important test of the supernovae feedback
modelling. Note that the scatter also increases at fainter magnitudes,
which is also observationally testable.

\begin{figure}
\centerline{
\psfig{file=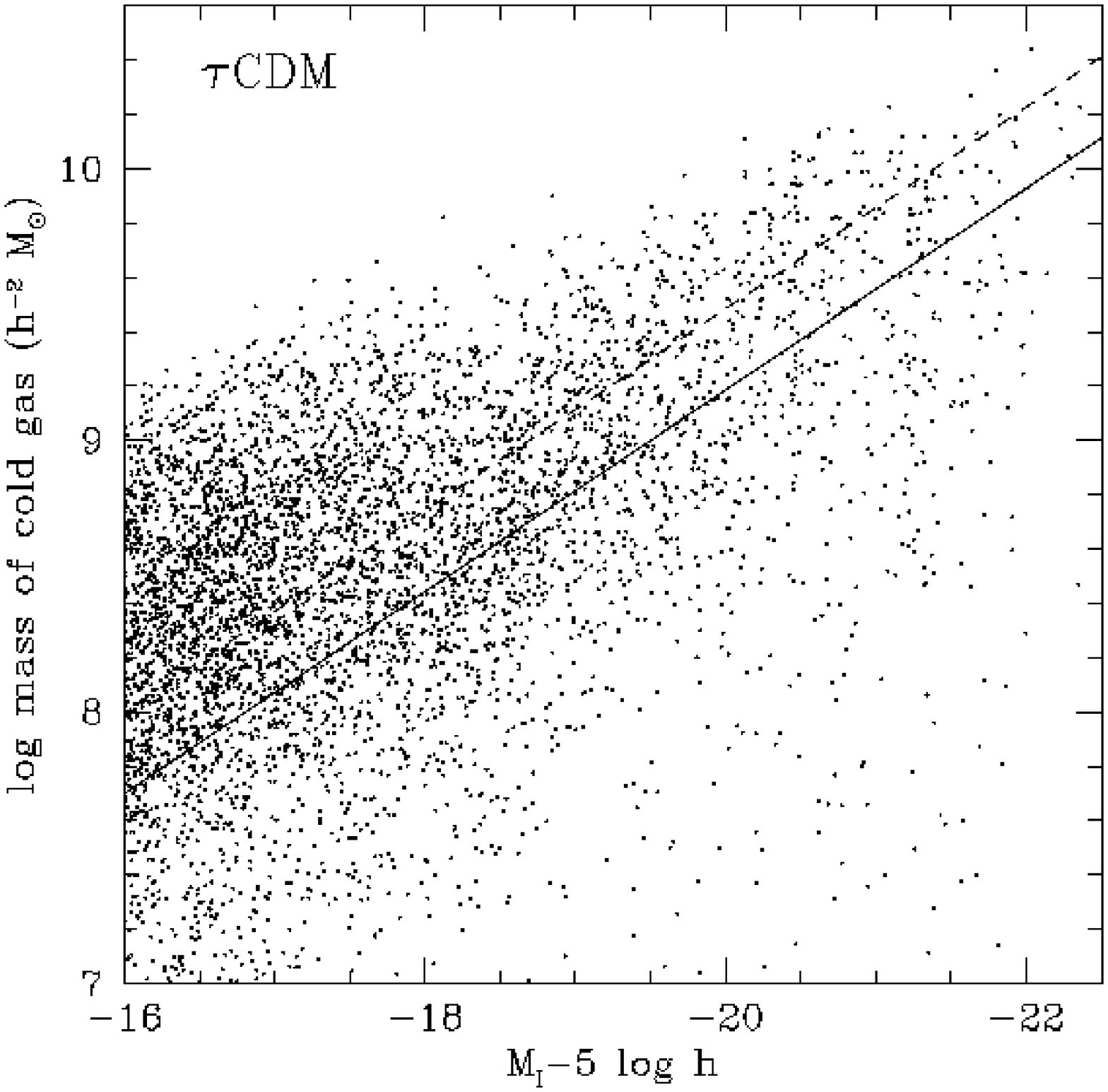,height=8.5truecm,width=8.5truecm}}
\centerline{
\psfig{file=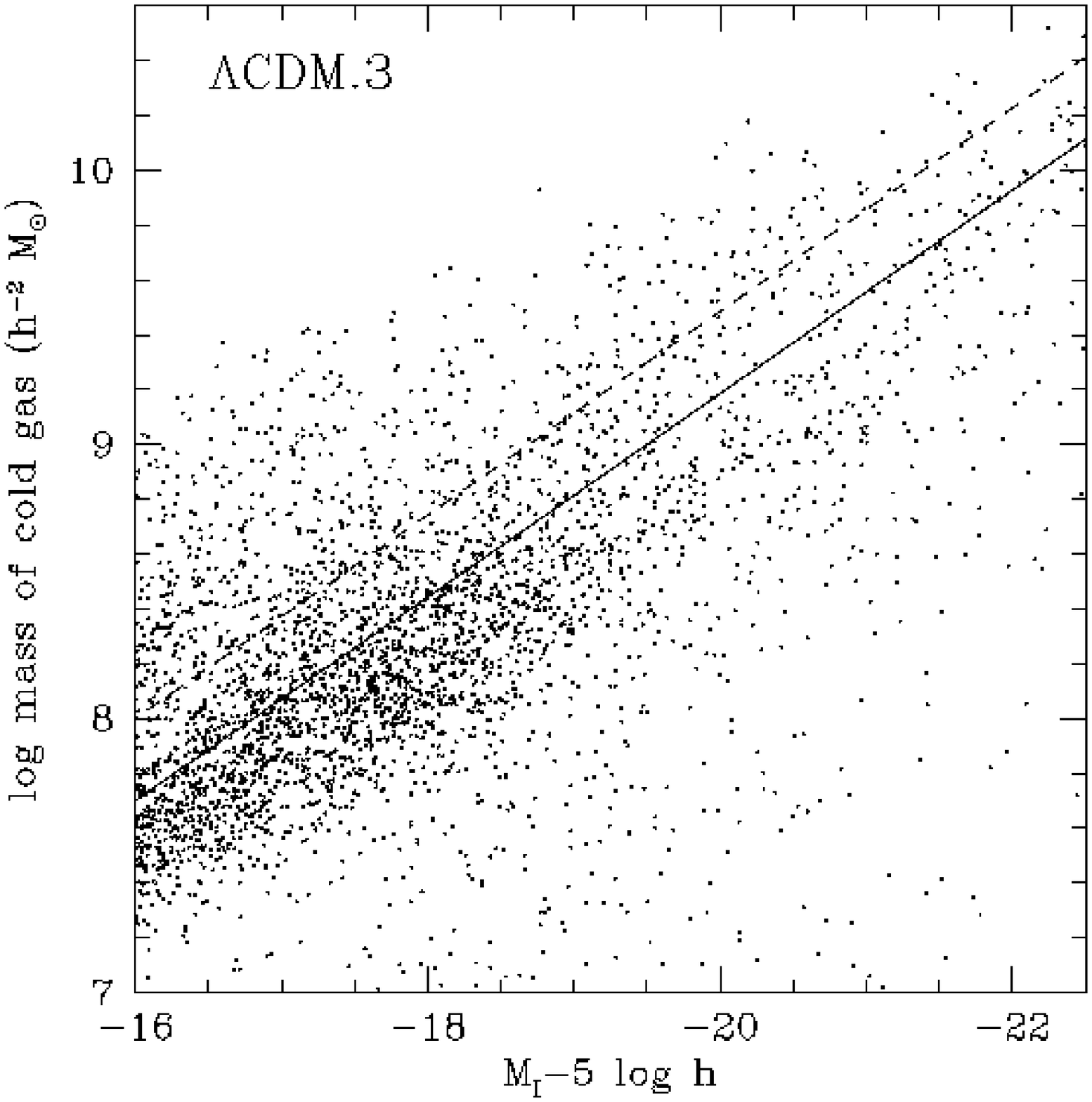,height=8.5truecm,width=8.5truecm}}
\caption{Small dots show the cold gas masses of the model galaxies 
from mock catalogs extracted from the New/Santa Cruz (fiducial)
\taucdm\ (top) and \lcdm3 (bottom) models. The solid line is an
approximate fit to ${\rm H_{I}}$ observations
\protect\cite{deblok:96b}. 
The dashed line is the same fit, with the gas masses multiplied by a
factor of two to allow for a contribution from cold gas not in the
form of $H_{\rm I}$. }
\label{fig:coldgas}
\end{figure}
It should be kept in mind that this comparison rests on the assignment
of model galaxy circular velocities as well as luminosities, and on
the conversion from circular velocity to linewidth. In the current
models we have assumed that all galaxies have perfectly flat rotation
curves out to the virial radius of the halo; i.e., that the circular
velocity measured by TF observations (typically at about two optical
disk scale lengths) is the same as the virial velocity of the dark
matter halo. This assumption clearly must break down in halos with
circular velocities larger than about 350-400 \kms, as no known
galaxies have rotation velocities this large. As we noted in
Section~\ref{sec:models:params}, if the profiles of dark matter halos
resemble the NFW profile, then $V_c$ at a few scale lengths will be
larger than at $r_{\rm vir}$ for smaller (galaxy) mass halos, and
smaller than at $r_{\rm vir}$ for larger (cluster) mass halos. In
addition, the dissipative infall of baryons will modify the inner
rotation curve \cite{blumenthal:86,flores:93}. Before attempting a
rigorous evaluation of the Tully-Fisher relation predicted by the
models, the halo and disk profiles should be modelled in more
detail. We intend to address this problem in future work.

\begin{figure}
\centerline{
\psfig{file=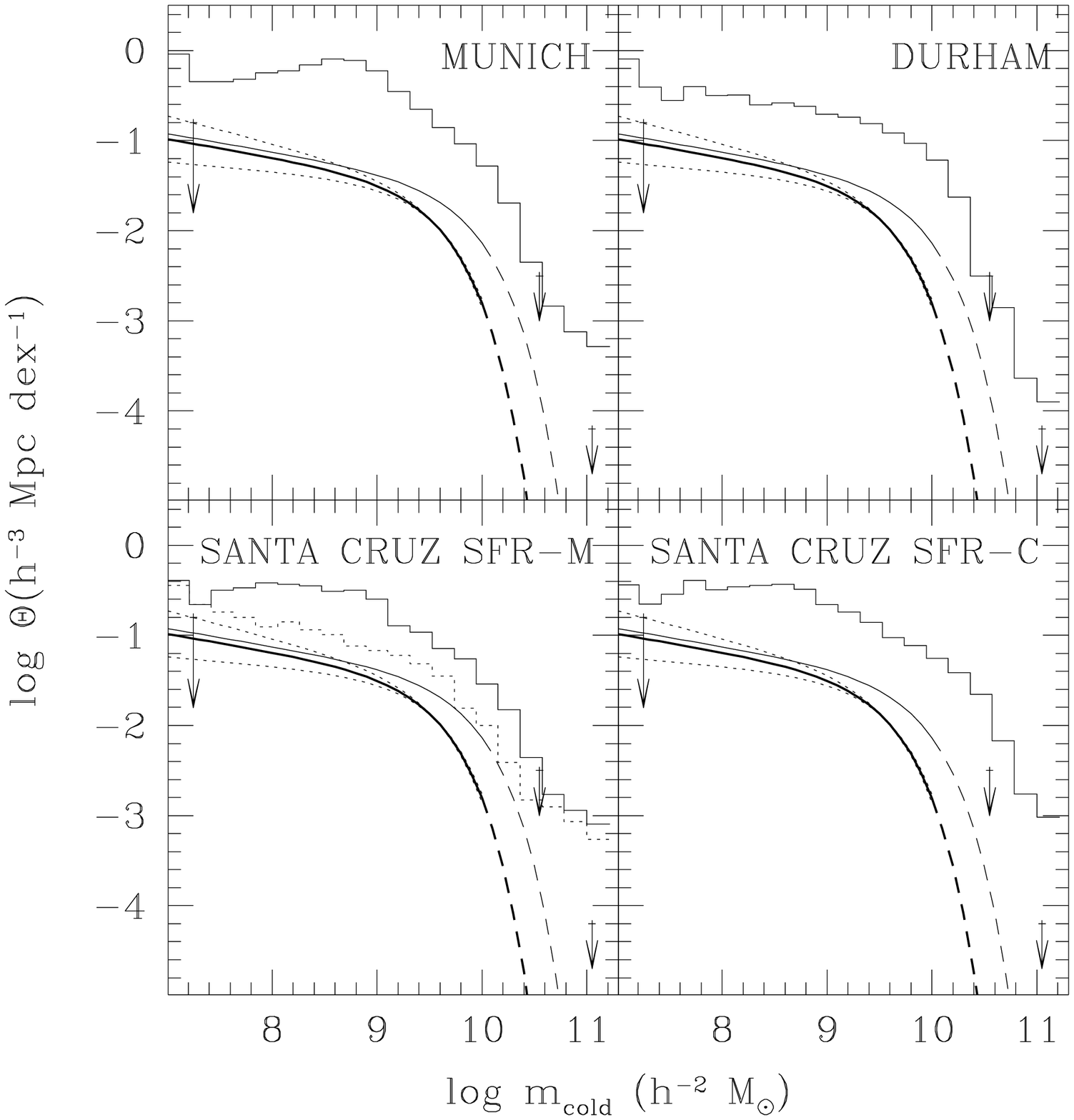,height=9truecm,width=8.9truecm}}
\centerline{
\psfig{file=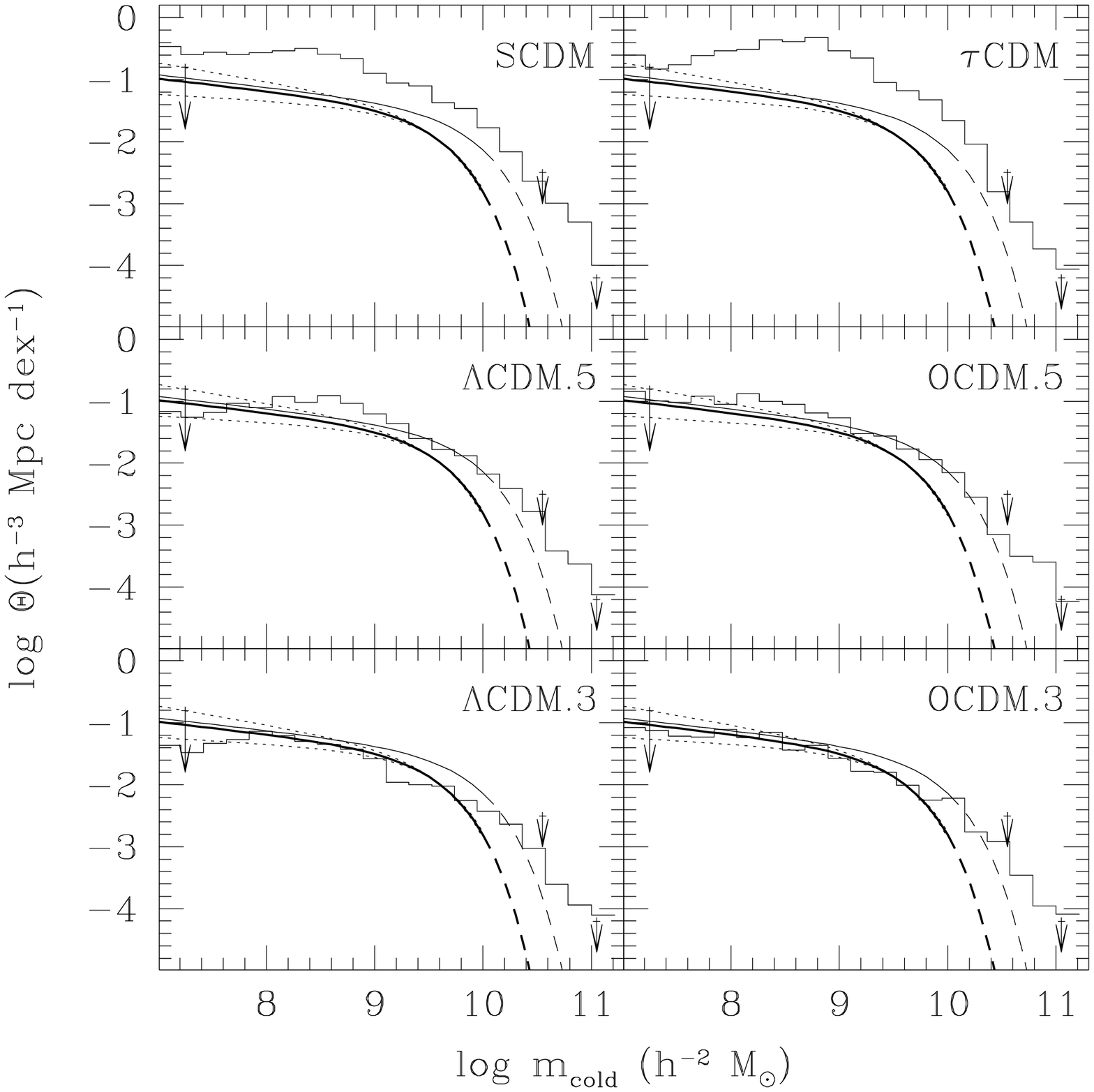,height=10truecm,width=8.9truecm}}
\caption{The $H_{\rm I}$ mass function. The two smooth lines show the Schechter
function fit to the results from \protect\citeN{zwaan}, where the
lower (bold) line is the actual fit and the upper line shows the
effect of increasing all the gas masses by a factor of two to account
for cold gas not in the form of $\rm H_{I}$. The survey is sensitive
in the range $10^7$ to $10^{10} \msun$, and the dashed lines show the
extrapolation of the Schechter function into the unprobed region. The
thin bold lines show the uncertainty in the faint end slope as given
by \protect\citeN{zwaan}. The arrows show upper limits from a
complementary Arecibo survey, also mentioned in
\protect\citeN{zwaan}. The histograms show the results from the models. 
The top panel shows the Classic/SCDM models with
different sf/fb packages, and the bottom panel shows the New/Santa
Cruz (fiducial) models for different cosmologies. }
\label{fig:coldgashist}
\end{figure}
\subsection{Cold Gas}
\label{sec:results:coldgas}
We now investigate the cold gas masses of galaxies in our models. This
is an important counterpart to studying the luminosities of
galaxies. Fig.~\ref{fig:coldgas} shows the mass of cold gas in the
model galaxies as a function of I magnitude for two examples of our
fiducial models (\taucdm\ and \lcdm3). The solid line shows an
approximate fit to local ${\rm H_{I}}$ data (see Fig.~12 of
\citeNP{deblok:96b}). Recall that we set the free parameters to match the
zero-point of this relation at $M_I-5 \log h \sim -21.8$, assuming
that the mass of ``cold gas'' in our model reference galaxy is
approximately a factor of two larger than the observed ${\rm H_{I}}$
mass to allow for molecular hydrogen (we neglect the additional
contribution of helium and ionized hydrogen). This corresponds to the
typical contribution of molecular hydrogen in an Sb-Sc type galaxy
\cite{young:89}. The observations show a large scatter, comparable to 
the scatter in the models. The models results are consistent with the
observed trend of gas mass with magnitude and the scatter in this
relation. It should be noted that we have not made any morphological
cuts on the model galaxies, whereas the observations are for late-type
galaxies. The results look similar for all the models.

We also investigate the ${\rm H_{I}}$ mass function, or the number
density of galaxies with a given ${\rm H_{I}}$ mass. This has been
estimated by the survey of \citeN{rao}, and more recently in the blind
${\rm H_{I}}$ survey described in \citeN{zwaan}. The latter should place
strong upper limits on the number of low surface brightness galaxies
(unless there is a very gas poor population) because it is not
optically selected. In Fig.~\ref{fig:coldgashist} we show the $H_{\rm
I}$ mass function for all the models, along with observations from
\citeN{zwaan}. All of the SCDM models show a considerable excess especially 
on the small-mass end. The \taucdm\ models show a somewhat smaller
excess, and the other models are in good agreement with the
observational limits across the range of ${\rm H_{I}}$ gas masses
probed by the observations.
\begin{figure}
\centerline{\psfig{file=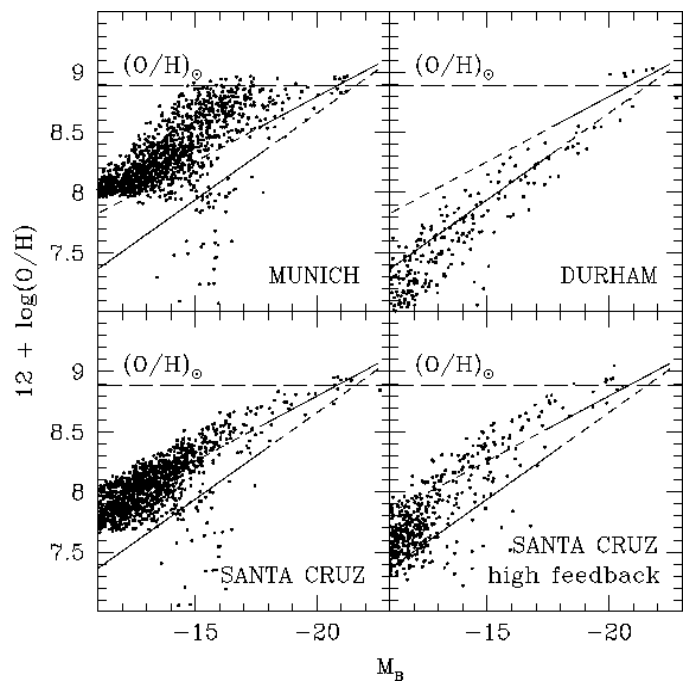,height=8.9truecm,width=8.9truecm}}
\caption{The metallicity-luminosity relation in the Classic/SCDM models 
(small dots), for galaxies within a ``local group'' (\refgal) sized
halo. Bold lines show fits to the observed relation for bright spirals
and for local dwarf galaxies (from the compilation in
\protect\citeNP{kobulnicky:98}). }
\label{fig:magmet_sf}
\end{figure}
\begin{figure}
\centerline{\psfig{file=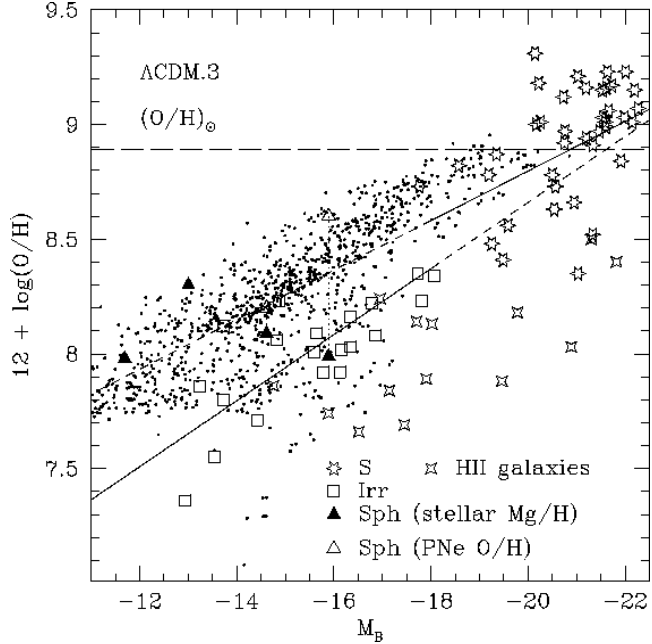,height=8.5truecm,width=8.5truecm}}
\caption{The metallicity-luminosity relation for a volume
limited mock catalog extracted from the New/Santa Cruz (fiducial) 
\lcdm3 models. Large symbols
show observations of bright spirals, dwarf irregulars and spheroidals,
and HII galaxies \protect\cite{kobulnicky:98}.}
\label{fig:magmet_lcdm3}
\end{figure}
\subsection{Metallicity-Luminosity Relation}
Nearby galaxies are known to exhibit a trend between their B-band
luminosities and their metal contents, in the sense that more luminous
galaxies are more metal-rich. The slope of the observed relation
derived for bright spirals ($M_B-5 \log h_{50} \la -18$) is shallower
than that for nearby dwarf galaxies
\cite{skillman:89,richer:95,zaritsky:94,kobulnicky:98}. 

Although we obtain a similar trend in the models (see
Fig.~\ref{fig:magmet_sf}), the detailed behaviour of the observations
is not well reproduced in any of the models. Recall that we have set
our yield parameter $y$ in order to obtain solar metallicity in our
approximately ``Milky Way'' sized reference galaxy. The relation that
we obtain depends on the treatment of metal and gas ejection by
supernovae. In the Munich package, none of the metal or gas is ejected
from the halo, and this package produces a very shallow relation with
a break at about $M_B=-15$. In the Durham package, all of the reheated
gas and metals are ejected from the halo, resulting in a very steep
relation even for the bright galaxies. In the Santa Cruz package, the
ejection of gas and metals is modelled using the disk-halo
approach. This leads to a relation which is consistent with the bright
galaxies, but dwarf galaxies that are too metal-rich compared to the
observations. This is the case even in the high feedback package. The
Durham package produces the best agreement with the observed
relation. However, it also produced an unacceptable degree of
curvature on the faint end of the Tully-Fisher relation.

\begin{figure}
\centerline{
\psfig{file=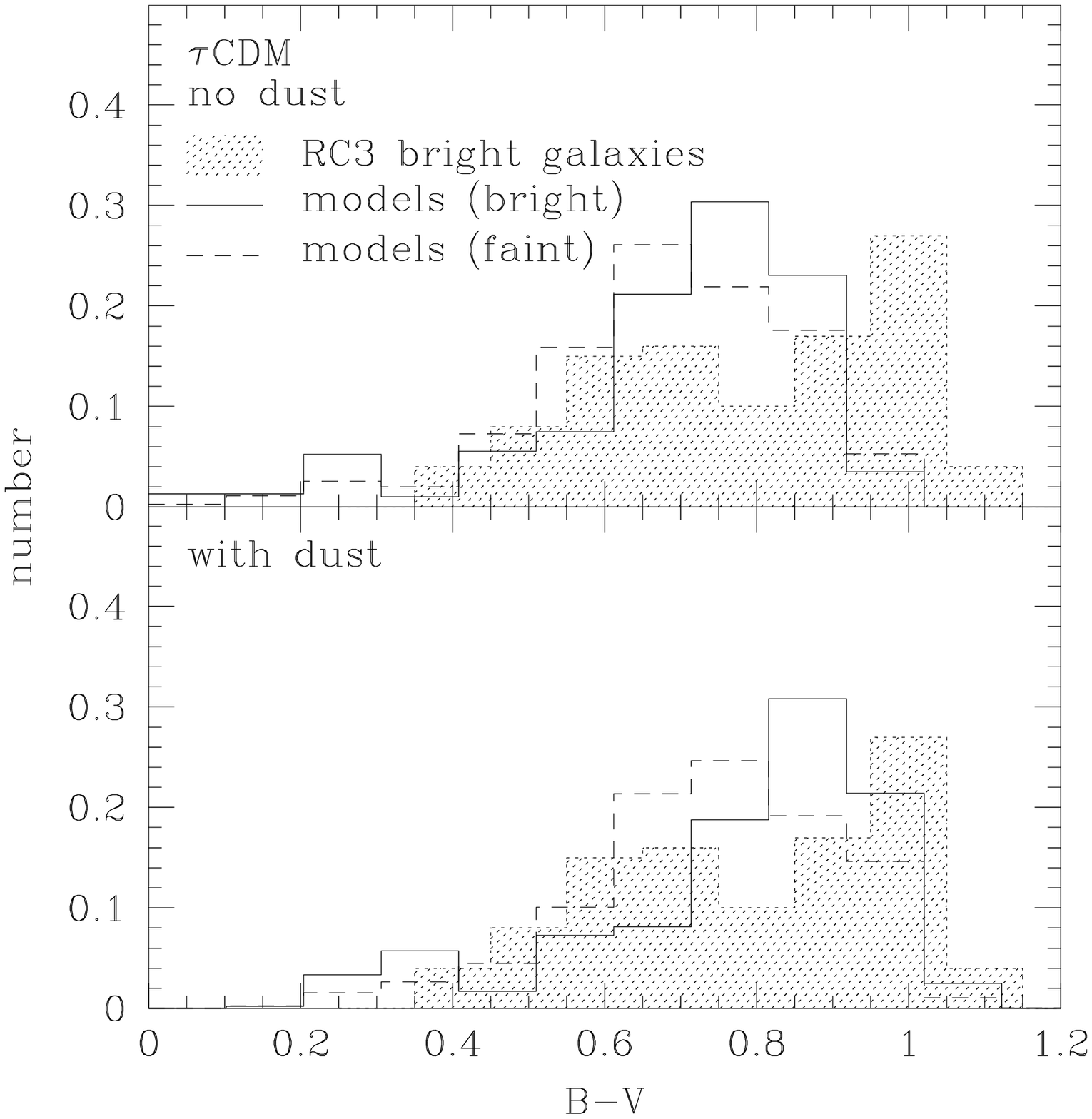,height=8.5truecm,width=8.5truecm}}
\centerline{
\psfig{file=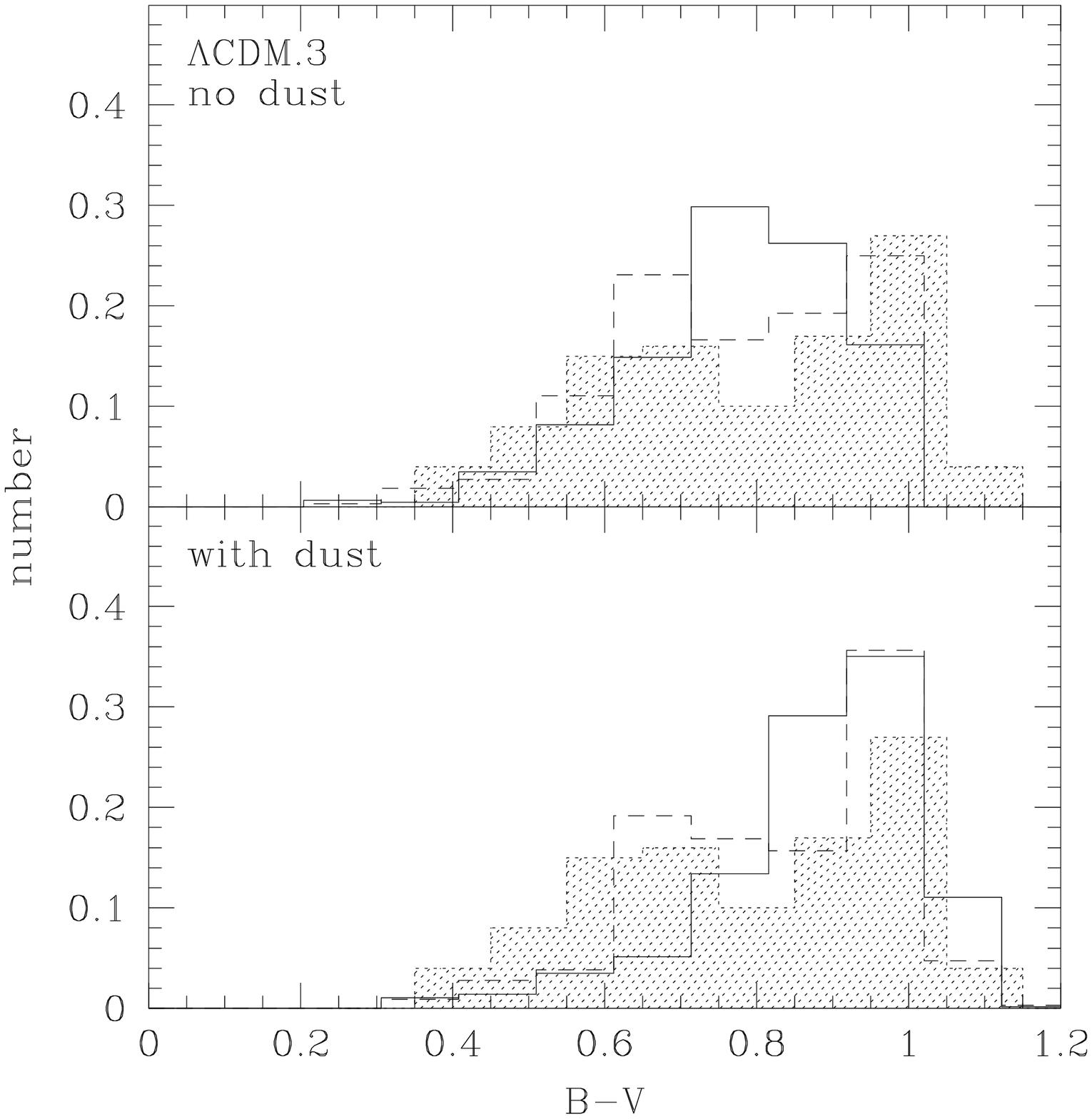,height=8.5truecm,width=8.5truecm}}
\caption{B-V colour histogram for galaxies in the New/Santa Cruz
(fiducial) models, in the \taucdm\ (top panel) and \lcdm3 (bottom
panel) cosmologies. The shaded histogram shows the observed colour
distribution of bright galaxies ($M_B-5\log h \ge -19.0$) from the RC3
catalog \protect\cite{rc3}. The unshaded histograms show the model
galaxies, selected to be brighter than $M_B-5\log h = -19.0$ (solid)
or $M_B-5\log h \ge -15.5$ (dashed). Within each panel, the top
half-panel shows the model results without the correction for dust
reddening, and the bottom half-panel shows the results with the
correction for dust. }
\label{fig:colors}
\end{figure}
The failure of all models in which ejected metals follow the ejected
gas may indicate that metal ejection is more efficient than gas
ejection.  This has been proposed on the basis of high-resolution
hydrodynamic simulations of dwarf galaxies
\cite{maclow-ferrara}. Our results support the strongly differential 
ejection efficiency proposed by \citeN{martin} on the basis of
observations. One should use caution in interpreting the observations,
however, because of possible systematic errors in the observational
determination of metal abundances in different types of galaxies and
using different methods \cite{kkp:98}. Interpreting the observations
is also complicated by the presence of metallicity gradients in large
galaxies, and the strong correlation of metallicity with surface
brightness \cite{garnett}. We postpone a more careful investigation of
these issues to future papers.

\subsection{Colours}
\label{sec:results:colours}
A familiar property of observed galaxies is the ``colour-magnitude''
relation: bright galaxies are observed to be redder than fainter
ones. It is an often repeated statement that hierarchical models of
galaxy formation generically predict that more massive objects form
``later'' than smaller mass objects. This statement is often
misinterpreted to imply that larger mass objects should be ``younger''
and therefore bluer than smaller ones. If the formation time of an
object is defined as the time when a given fraction of its mass has
been assembled into a single progenitor, then it is true that larger
mass objects have later formation times than smaller mass
objects. However, if we define ``age'' as the time spent in the sort
of environment where we expect that star formation is able to occur
(i.e., within a collapsed halo), then the mean age of the material in
large mass halos is older than in smaller mass halos. This is because
large mass halos are associated with higher peaks in the density
field, which collapse earlier.

In the SAMs, we find that when the effects of dust and metallicity are
neglected, we obtain flat colour-magnitude relations or (depending on
the model and the colour bands in question) bright galaxies that are
only slightly redder (see Fig.~\ref{fig:colors}; top panel) than the
faint population. The bright galaxies are also a bit too blue overall
compared with observations (the observed colour distribution of bright
galaxies in the RC3 catalog \cite{rc3} is shown for
comparison). Inclusion of dust extinction shifts the colour
distribution towards the red, and shifts the bright galaxies more than
the faint ones because of the differential nature of our dust
recipe. This brings the optical colours into fairly good agreement
with the observations. In addition,
\citeN{kauffmann:cm} have shown that the inclusion of metallicity
effects on the model spectra (here we have used only the solar
metallicity stellar population models) can also produce the observed
colour-magnitude slope. Presumably the observed trend is a combination 
of these two effects. 

\begin{figure}
\centerline{\psfig{file=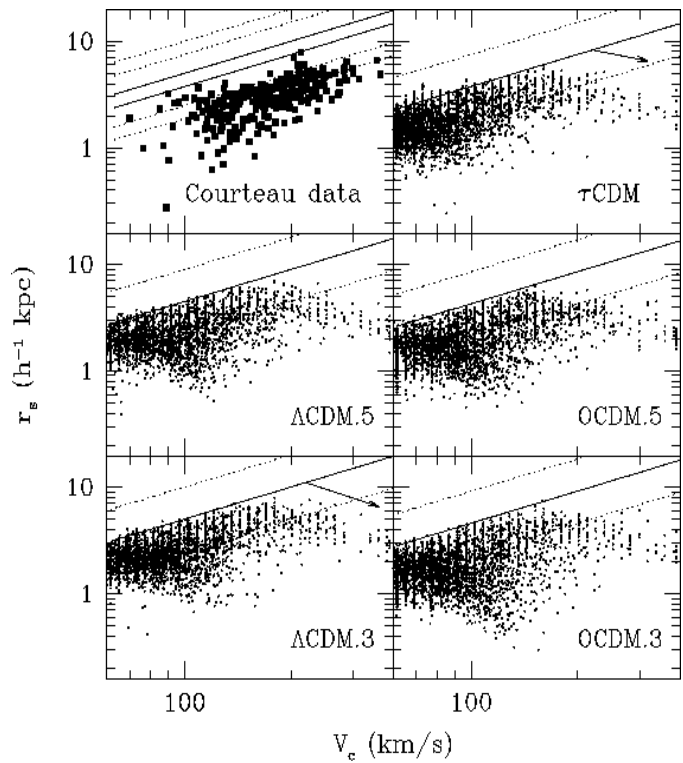,height=10truecm,width=8.9truecm}}
\caption{Exponential scale radius vs. circular velocity. The large filled
squares (top left panel) show the observations of
\protect\citeN{courteau}. The small dots show the results of the
fiducial Santa Cruz models in various cosmologies as indicated. Bold
solid lines show the relation we would obtain if all gas fell in from
the virial radius of the halo, for $\lambda_{H}=0.05$ (50 percent
point). Bold dashed lines show the same relation for
$\lambda_{H}=0.025$ (10 percent) and $\lambda_{H}=0.1$ (90
percent). These vary slightly depending on cosmology. The \taucdm\
(lower) and \lcdm3 (upper) relations are shown on the upper left panel 
with the observations as a reference point (other models are
intermediate). The arrows on the \taucdm\ and \lcdm3 panels show the
corrections to this relation predicted from the fitting formulae of
\protect\citeN{mmw:98} (see text). }
\label{fig:sizes}
\end{figure}
\subsection{Sizes}
\label{sec:results:sizes}
We estimate the exponential scale radii of our model disks using the
approach described in Section~\ref{sec:models:sizes}. We show the
relationship between scale radius and circular velocity in
Fig.~\ref{fig:sizes}. The upper left-most panel shows the observations
of \citeN{courteau} for late-type galaxies, where $V_c$ is the disk
rotation velocity at 2.2 scale-lengths, and $r_s$ is the exponential
scale length in the $r$ band. In the models, the scale radius that we
estimate represents the total baryonic mass (stars and cold gas) in
the disk. The scale radius of the stellar mass may be smaller if star
formation is more efficient in the inner parts of the disk, and the
scale of the optical light may be smaller yet. The bold solid lines
indicate the relation $r_s = 1/\sqrt{2} \lambda_{H} r_{\rm vir}$,
which we would obtain if the gas fell in from the virial radius of the
halo. As we have discussed in Section~\ref{sec:models:sizes}, we have
used a constant value of $\lambda_{H}=0.05$ for all halos, so the only
source of scatter in the relation that we obtain is from the different
cooling radii of the halos, which occur as a result of the scatter in
halo merging histories (see Figure~\ref{fig:cooling}). The SAM dots
always lie at smaller $r_s$ than the bold lines because the cooling
radius is always smaller than the virial radius. The bold dashed lines
show the same relation for $\lambda_{H}=0.1$ and $\lambda_{H}=0.025$,
which are the 10 and 90 percent points of the distribution of
$\lambda_{H}$ used by \citeN{mmw:98}.

As we discussed in Section~\ref{sec:models:sizes}, we have not
included the ``back reaction'' of baryons on the dark matter during
their collapse, which will tend to lead to smaller scale radii
\cite{blumenthal:86,flores:93,mmw:98} and will also modify the
rotation curve. The use of a more realistic halo profile (e.g. NFW)
will also change these results. The correction due to these effects,
as predicted by the fitting formulae of \citeN{mmw:98}, is shown by
the arrows in the \taucdm\ and \lcdm3 panels (the correction is
relatively insensitive to cosmology, and is similar for the other
models). The direction of the correction is to produce galaxies with
\emph{larger} $V_{\rm max}$ and \emph{smaller} scale radii.
The details of these corrections depend on the assumed halo profile
and disk baryon fraction as well as other parameters. We intend to
incorporate improved modelling of disk sizes and rotation curves into
our models and present more detailed predictions of these quantities
in the near future. 

In the meantime, several things are worth noting. An obvious
difference in the current model predictions is the break in the
$r_s$-$V_c$ relation at about 200 \kms: the trend reverses and the
scale radii start to decrease at larger $V_c$. This is caused by the
decreasing cooling radius in halos with higher virial temperature (see
Fig.~\ref{fig:cooling}). No such break is evident in the observations,
although the sample contains a relatively small number of galaxies
with $V_c > 200$
\kms. The rightward shift indicated by the arrows (due to ``peaking
up'' of the rotation curve caused by the effects mentioned above) may
solve this problem, but will also make the disks too small at a given
$V_c$ compared to the Courteau data. Most of the effects we have
mentioned indicate that we may already be systematically
over-estimating the disk sizes. We therefore may be facing a puzzle
similar to the ``angular momentum'' problem found in N-body
simulations with hydrodynamics (i.e., disks are too small and
concentrated at a given circular velocity compared to observations,
cf. \citeNP{steinmetz:99}).

\section{Summary and Discussion}
\label{sec:summary}
We have presented new semi-analytic models of galaxy formation and
shown that these models can reproduce many key observational
properties of galaxies in the local Universe. Our approach is similar
to that introduced by \citeN{kwg} and \citeN{cafnz}, but we have
introduced several modified or new ingredients, including:
\begin{itemize}
\item Somerville-Kolatt method for ``planting'' merger trees (\citeNP{sk}; 
shown to give good agreement with merger trees extracted from N-body 
simulations in \citeNP{slkd}).
\item Improved Sheth-Tormen model \cite{sheth-tormen} for the mass function 
of dark matter halos (an improved version of the Press-Schechter model
which gives much better agreement with the reusults of N-body
simulations).
\item ``Dynamic halo'' model for gas cooling (includes the effects 
of halo merger events on the density and temperature of the hot halo gas).
\item ``Disk-halo'' model for supernovae feedback (models the ejection of 
cold gas from the disk and global (dark matter) potential seperately)
\item Dust extinction based on the empirical recipe of \citeN{wh}
\item Galaxy mergers due to satellite collisions, using the simulation-based 
approximation of \citeN{makino-hut}
\item More detail modelling of starbursts based on hydrodynamical 
simulations \cite{mihos:94,mihos:96}
\end{itemize}
We have investigated several different ``packages'' of recipes for
star formation and supernovae feedback in order to gain better
understanding of the importance of the way in which these processes
are parameterized. We have also illustrated the results of varying the
most important free parameters in our models.

We have addressed the long-standing problem of the physical
explanation of the observed luminosity function within CDM-type
hierarchial models of structure formation. Early in the history of
CDM, it was noted that the mass function of dark matter halos, whether
predicted by analytic models like Press-Schechter or derived from
N-body simulations, has a steep power-law slope ($\alpha\sim-2$) for
masses $\la 10^{14}
\hmsun$, and an exponential cut-off at $\sim 10^{14} \hmsun$, much
larger than the expected mass of the halos surrounding $L_*$ galaxies,
as estimated from their internal velocity dispersions. It was proposed
that feedback due to supernovae could suppress star formation in small
mass halos, leading to a flatter faint end slope, and inefficient gas
cooling in large mass halos could cause the ``knee'' at
$L_*$. However, the first generation of SAMs, which attempted to
actually model these processes in some detail, encountered some
difficulties. The Munich models produced the correct (B-band) TFR
slope, but the faint-end slope of the luminosity function was still
too steep. In addition, unless an ad-hoc cutoff was applied, in which
gas cooling was turned off by hand in halos larger than 500
\kms, these models did not produce a ``knee'' in the
luminosity function and showed an excess of very bright
galaxies. These models were normalized to the observed luminosity of
the Milky Way Galaxy, and were claimed to reproduce the observed
zeropoint of the local B-band TFR. The Durham models produced a
luminosity function with a ``knee'', and the free parameters were
adjusted in order to match its location with that of the observed
B-band luminosity function. Their luminosity functions showed a
flatter faint-end slope, in better agreement with observations, but
produced a TFR with a zeropoint offset of about 2 magnitudes and a
serious deviation from the observed power-law behavior on the small
$V_c$ end.

We have clarified the reasons for some of these differences. First, we
have explained how it is that, although the Munich group claimed to
reproduce the observed zeropoint of the TFR, in fact their model
galaxies were $\sim 2$ magnitudes too faint at a given circular
velocity compared to the TFR derived from recent large I-band
samples. Using our models with sf/fb recipes chosen to be similar to
those of the Munich and Durham group, we showed that when the models
are normalized in the same way, the luminosity function and TFR are
nearly identical for \emph{bright/large $V_c$} galaxies ($M_{B}- 5\log
h \la -20$, $V_c \ga 220$ \kms). The results differ substantially only
for faint/small $V_c$ galaxies, and this difference can be traced
mainly to the stronger supernovae feedback recipe assumed by the
Durham group. We demonstrate this by showing that as we turn up the
parameter that represents the fraction of supernovae energy deposited
in the cold gas ($\epsilon_{SN}^0$), our results move continuously
from a situation resembling the Munich models (steep LF, powerlaw TFR)
to one resembling the Durham models (flat LF, curved TFR). This works
because in our ``disk-halo'' feedback model, the parameter
$\epsilon_{SN}^0$ affects not only the total amount of gas that is
reheated, but also the fraction that is ejected from the halo. We
issue a warning, however, that although the supernovae feedback
efficiency is the dominant factor determining these results at $z=0$,
the redshift evolution is also very sensitive to the assumed star
formation recipe. This will be illustrated in detail in a companion
paper \cite{spf}.

\begin{figure}
\centerline{\psfig{file=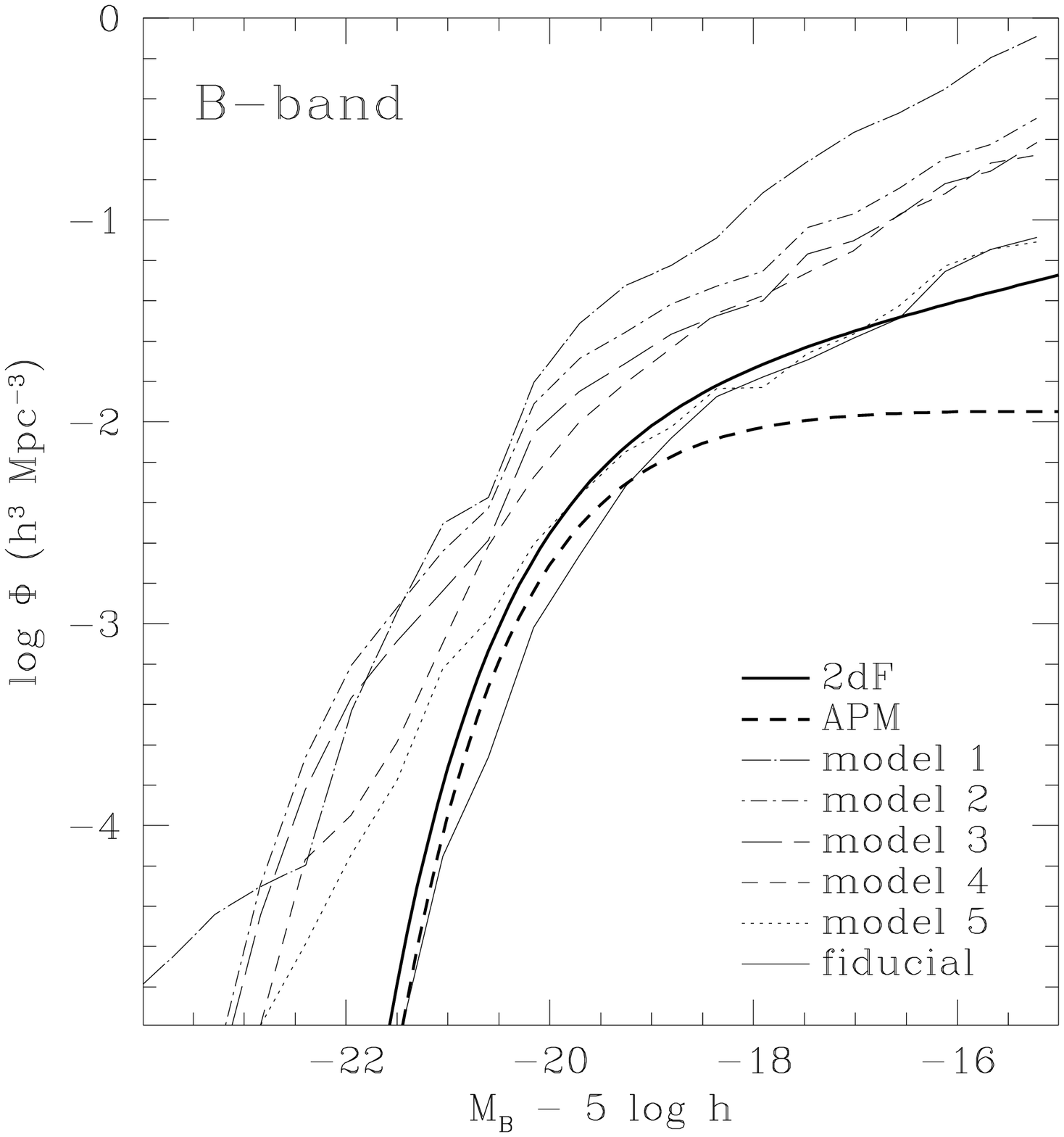,height=8.5truecm,width=8.5truecm}}
\centerline{\psfig{file=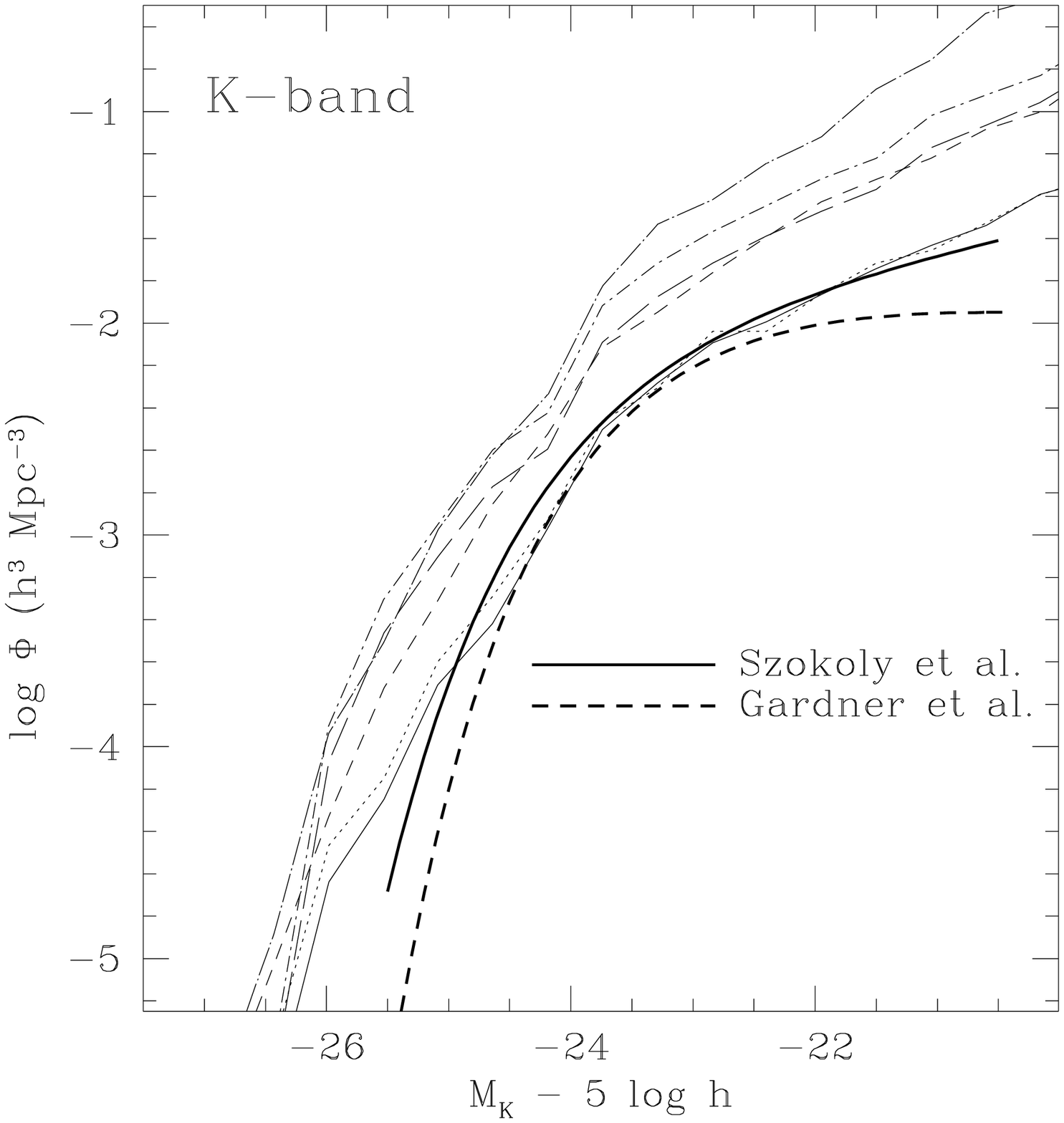,height=8.5truecm,width=8.5truecm}}
\caption{The effect of various model variations on the model B and K-band 
luminosity functions, introduced one by one. The new feature of each
model is in italics. 
{\bf Model 1:} Munich-style feedback, original
Press-Schechter mass function, static halo cooling, SCDM, no dust
correction. 
{\bf Model 2:}
\emph{disk-halo feedback}, original Press-Schechter mass function, static
halo cooling, SCDM, no dust correction. 
{\bf Model 3:} disk-halo feedback,
\emph{Sheth-Tormen mass function}, static halo cooling, SCDM, no dust
correction. 
{\bf Model 4:} disk-halo feedback, Sheth-Tormen mass function,
\emph{dynamic halo cooling}, SCDM, no dust correction. 
{\bf Model 5:} disk-halo
feedback, Sheth-Tormen mass function, dynamic halo cooling,
\emph{$\Lambda$CDM.5}, no dust correction. 
{\bf Fiducial model:} disk-halo feedback, Sheth-Tormen mass function,
dynamic halo cooling, $\Lambda$CDM.5, \emph{with dust correction}. In
the top panel, the bold lines show the observed B-band luminosity
function from the 2dF survey (higher line) and the APM survey (lower
line). In the bottom panel, the bold lines show the observed K-band
luminosity function from \protect\citeN{szokoly:98} and
\protect\citeN{gardner:97}. }
\label{fig:lf_compare}
\end{figure}
We have shown that the observed B and K band luminosity functions and
the Tully-Fisher relation can be reproduced simultaneously in our
models. This improvement is not due to any one effect but is the
result of many combined factors, as summarized below. We first
normalize our models to fix the zeropoint of the TFR for a typical
$L_{*}$ galaxy. This requires us to use a stellar mass-to-light ratio
approximately a factor 2 to 3 times higher than the published Munich
and Durham models, but corresponds to taking the predictions of the
Bruzual-Charlot stellar population models at close to face value and
is in better agreement with observational estimates. If we then use
cooling, feedback, and star formation recipes similar to the published
Munich models, we obtain a B-band luminosity function with several
problems: the overall normalization is too high, the faint end slope
is too steep, and there is a ``tail'' of bright galaxies (see
Fig.~\ref{fig:lf_compare}, model 1). Fig.~\ref{fig:lf_compare} shows
the effect of introducing various changes, one by one, which
eventually lead to our fiducial model choices. These are summarized
below. We remark whether the effect is important on the bright ($M_B-5
\log h
\sim -21.5$) or faint ($M_B-5 \log h \sim -16.5$) end of the
luminosity function, and by what factor the B-band luminosity function
changes at this magnitude. These factors should be considered
approximate only, and may be read from the top panel of
Fig.~\ref{fig:lf_compare}.
\begin{itemize}
\item ``disk-halo'' feedback model (faint end, factor of 2.5)
\item Sheth-Tormen mass function (overall, factor of 1.5)
\item ``dynamic halo'' cooling model (bright end, factor of 3)
\item low $\Omega \sim 0.5$ cosmology (bright end, factor of 1.6; 
faint end, factor of 3)
\item dust extinction (bright end, factor of 16)
\end{itemize}
Note that the observations have also changed --- the solid bold curve
in Fig.~\ref{fig:lf_compare} shows the luminosity function derived
from the recent 2dF survey
\cite{twodf}, which is in excellent agreement with the LF from the
deep ESO slice \cite{zucca}. Both have considerably steeper faint-end
slopes than the LF derived from the APM survey \citeN{loveday}, which
was the standard at the time of much of the earlier modelling. The
more recent observations are easier to reconcile with the models.
A similar accounting may be done for the K-band luminosity function
(Fig.~\ref{fig:lf_compare}, bottom panel). 

We therefore conclude that the very strong feedback and suppression of
star formation in small $V_c$ galaxies assumed in the Durham models is
not necessary in order to reproduce the observed luminosity function,
and is disfavored as it produces curvature on the small $V_c$ end of
the TFR.

In our $\Omega_0=1$ models (SCDM and \taucdm), we find that in order
to produce galaxies with large enough luminosities and gas masses, we
must assume values of the baryon fraction ($f_{\rm baryon}
\sim 0.1-0.12$) which are rather high 
compared with estimates from observations of high-redshift deuterium
\cite{tytler:99}, though consistent with estimated baryon fractions in 
groups and clusters. In our best $\Omega_0=1$ models, we find good
agreement with the general
\emph{shape} of the LF, but the model LF is too high by an overall
factor of $\sim 3$ in B and 2.5 in K. The mass function of cold gas is
also a factor of $\sim 5$ higher than estimates from blind ${\rm
H_{I}}$ surveys. In order to reconcile these models with observations,
we would have to believe that there is a substantial population of
galaxies, including some with large total masses, that are undetected
in optical emission or radio emission from cold ${\rm H_{I}}$
gas. This seems unlikely, though not impossible. Thus, although we
cannot say that $\Omega_0=1$ is ruled out, our results are certainly
more easily compatible with models in which $\Omega_0 \sim 0.3-0.5$,
with or without a cosmological constant.

The same fiducial models produce good agreement with observations of
the mass function of cold ${\rm H_{I}}$ gas and the magnitude-${\rm
H_{I}}$-mass relation. When we normalize our models to produce a
``Milky-Way'' galaxy with solar metallicity, the metallicities of
dwarf galaxies in our models are somewhat higher than the average
metallicity of nearby dwarf galaxies, i.e. the slope of the
metallicity-luminosity relation is too shallow. This may be evidence
that metals are ejected by supernovae more efficiently than the cold
gas, or an indication that our ``constant yield'' approach to
modelling chemical evolution is too simplistic. Alternatively, it may
be due to systematic uncertainties in deriving observational estimates
of metal abundances in different types of galaxies \cite{kkp:98}. Our
fiducial models produce good qualitative agreement with the optical
colors of bright galaxies, and reproduce the observed color-magnitude
trend, when dust extinction is included. Although the relationship
between the exponential scale radius and circular velocity of disks
that we estimate is in reasonably good qualitative agreement with
observations, we conclude that more detailed modelling is necessary.

A great strength of the SAM technique is that one can make
self-consistent predictions pertaining to a wide variety of
observations. In companion papers, we investigate our predictions for
the properties of high-redshift galaxies and the history of stars,
cold gas, and metals at high redshift \cite{spf}, and extend our
predictions for local galaxies to shorter (far UV) and longer (far IR
to sub-mm) wavelengths \cite{sbmp,bsp}.

\section*{Acknowledgements} 
We would like to thank Carlton Baugh, Shaun Cole, Carlos Frenk, Cedric Lacey,
Simon White, and especially Guinevere Kauffmann for useful discussions and for
answering questions about their models. We have also benefited greatly from
discussions and collaboration with Sandra Faber, Avishai Dekel, Tsafrir Kolatt,
Donn MacMinn, James Bullock, Ari Maller, Georg Larsen and Mike Fall. We thank
Stephane Courteau and Chip Kobulnicky for making their observational data
available to us electronically, and Stephane Charlot for help with the GISSEL
models. We thank the anonymous referee for suggestions which helped to improve
this manuscript. RSS acknowledges support from a GAANN fellowship at UCSC, and
a President's fellowship at the Hebrew University. This work was also supported
by grants from NSF and NASA at UCSC and the BSF at HU.

\appendix
\section{Spherical Collapse in a General Cosmology}
\label{app:scollapse}
\begin{figure}
\centerline{\psfig{file=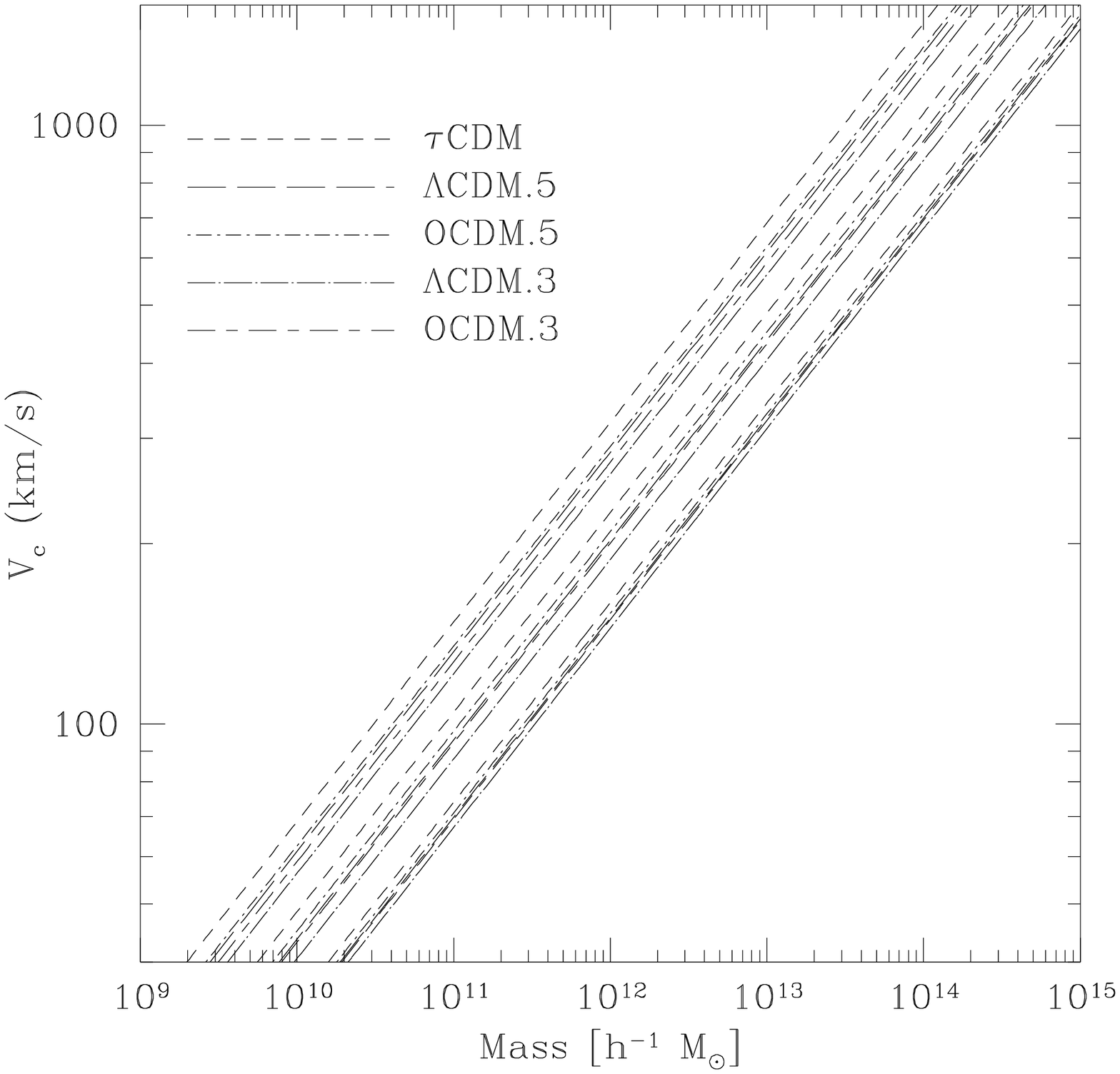,height=8.5truecm,width=8.5truecm}}
\caption{The relationship between halo mass and virial velocity from the 
spherical tophat model, at $z=0$ (bottom set of lines), $z=1$, and
$z=3$ (top), for the cosmologies discussed in the text. The relation
depends (weakly) on cosmology and (strongly) on redshift.}
\label{fig:massvc}
\end{figure}
We need to be able to relate the mass, radius, and velocity dispersion
of dark matter halos for any given redshift. This is made possible
using the spherical collapse model, one of the apparently gross
oversimplifications that seems to work surprisingly well. We imagine a
spherical patch of the universe with a uniform overdensity $\delta_i$
within a radius $r_i$ at a very early time $t_i$ (often called a
``top-hat'' perturbation). We assume that the collapsing shells of
matter do not cross. If we consider a particle at radius $r$,
Birkhoff's theorem \cite{birkhoff} tells us that we can ignore the
mass outside this radius in computing the motion of the particle. The
equation of motion for our particle (in physical, rather than
comoving, coordinates) is then
\begin{equation}
\frac{d^2r}{dt^2} = -\frac{GM}{r^2} + \frac{\Lambda}{3} r
\end{equation}
where $M = (4\pi/3)r_i^3\rho_b(t_i)(1+\delta_i)$ and $\rho_b(t_i)$ is the
background density of the universe at $t_i$. Integrating this equation gives
\begin{equation}
\dot{r} = H_0\left[\frac{\Omega_0}{r}(1+\delta_i)\frac{r_i^3}{a_i^3} 
+ \Omega_{\Lambda}r^2 - K \right]
\end{equation}
where $K$ is a constant of integration. We may fix this by noting that if we
have picked $t_i$ early enough that $\Omega \sim 1$ at that time, linear theory
tells us that the initial velocity is
\begin{equation}
\label{eqn:pert}
\dot{r}(t_i) = H_0 r_i \left(1-\frac{\delta_i}{3} \right)\sqrt{\frac{\Omega_0}{a_i^3} 
+ \frac{\Omega_R}{a_i^2} + \Omega_{\Lambda}}.
\end{equation}
\cite{peebles:1984}. At the point of maximum expansion, or ``turnaround'',
$\dot{r} = 0$. If we set equation (\ref{eqn:pert}) to zero, we obtain
a cubic equation for $r_{ta}$, the radius of the perturbation at
turnaround, which must be solved numerically for the general cosmology
given here, but for special cases it can be solved analytically
(cf. \citeNP{pad}). From a symmetry argument, we note that the time
when the perturbation collapses to a point, $t_{\rm coll}$, is always
twice $t_{ta}$ (the time at maximum expansion). We can now write an
implicit equation for the mass of a perturbation that is collapsing at
$t_{\rm coll}$:
\begin{equation}
t_{\rm coll} = 2 \int_{0}^{r_{ta}} \frac{dr}{\dot{r}} \, .
\end{equation}
We know the mass and the radius at turnaround, so we can calculate the density
of the perturbation at turnaround, $\rho_{ta}$.

Of course the perturbation will not really collapse to a point. Before that
happens, shell crossing will occur, and it will virialize. We can find the
radius after viralization in terms of the turnaround radius using the virial
theorem. The total energy at turnaround is \cite{lahav:1991}
\begin{equation}
E = U_{G, ta} + U_{\Lambda, ta} = -\frac{3}{5}\frac{GM^2}{r_{ta}} -
\frac{1}{10}\Lambda M r_{ta}^2
\end{equation}
where the second term is due to the cosmological constant. Now using
the virial theorem for the final state:
\begin{equation}
T_{f} = -\frac{1}{2}U_{G,f} + U_{\Lambda, f} \, .
\end{equation}
From conservation of energy we then have $\frac{1}{2}U_{G,f} +
2U_{\Lambda,f} = U_{G,ta} + U_{\Lambda, ta}$. This leads to a cubic
equation for the ratio of the virial radius $r_{\rm vir}$ to the
turnaround radius $r_{ta}$. We now know $r_{\rm vir}$ and can write down
the virial density 
\begin{equation}
\Delta_c(z) \equiv \frac{\rho_{\rm vir}
\Omega(z)}{\Omega_{0}\rho_{c}^{0}(1+z)^3}. 
\end{equation}

We now have a relationship between the mass, virial radius, and
collapse redshift $z$. If we assume a radial profile for the
virialized halo, we can use the virial theorem again to relate these
quantities to the velocity dispersion. If we assume that the halo is a
singular isothermal sphere, $\rho
\propto r^{-2}$, truncated at the virial radius, then we have
\begin{equation}
\frac{3}{2} \sigma^2 = \frac{GM}{2r_{\rm vir}} - \frac{\Lambda r_{\rm vir}^2}{18}
\end{equation}
or, in terms of the circular velocity $V_c$, assuming $V_c^2 = 2\sigma^2$:
\begin{equation}
\label{eqn:virvelocity}
V_{c}^{2} = \frac{GM}{r_{\rm vir}} - \frac{\Omega_{\Lambda}}{3}H_0^2 r_{\rm vir}^2
\end{equation}
We can now translate between mass and velocity dispersion at any given
redshift. Note that in universes with a non-zero cosmological
constant, halos of a given circular velocity are less massive because
of the $\Lambda$ contribution to the energy.

In practice, we use the fitting formula of \citeN{bryan:97} for the virial
density:
\begin{equation}
\Delta_c = 18 \pi^2 + 82x -39 x^2 
\end{equation}
for a flat universe and 
\begin{equation}
\Delta_c = 18 \pi^2 + 60x -32 x^2 
\end{equation}
for an open universe, where $x\equiv \Omega(z)-1$. This formula is accurate to
1\% in the range $0.1 \le \Omega \le 1$, which is more than adequate for our
purposes. We now can write down the general expression for $r_{\rm vir}$ in
closed form.
\begin{equation}
r_{\rm vir} = \left[\frac{M}{4\pi} 
\frac{\Omega(z)}{\Delta_c(z)\Omega_0\rho_{c,0}}\right]^{1/3}
\frac{1}{1+z}.
\end{equation}
In conjunction with equation~\ref{eqn:virvelocity}, this allows us to calculate
the circular velocity and viral radius for a halo with a given mass at any
redshift $z$. These expressions are valid for open cosmologies with $\Lambda=0$
and flat cosmologies with non-zero $\Lambda$. 


\bibliographystyle{mnras}
\bibliography{mnrasmnemonic,sp2}

\end{document}